\documentclass[conference]{IEEEtran}
\IEEEoverridecommandlockouts
\usepackage{cite}
\usepackage{amsmath,amssymb,amsfonts}
\usepackage{algorithmic}
\usepackage{graphicx}
\usepackage{textcomp}
\usepackage{xcolor}
\usepackage{tabularx}
\usepackage{hyperref}
\usepackage{multirow}
\usepackage{longtable}
\usepackage{url}
\usepackage{breakurl}

\def\BibTeX{{\rm B\kern-.05em{\sc i\kern-.025em b}\kern-.08em
    T\kern-.1667em\lower.7ex\hbox{E}\kern-.125emX}}
\begin{document}

\title{Emotions in the Loop: A Survey of Affective Computing for Emotional Support}

\author{\IEEEauthorblockN{Karishma Hegde}
\IEEEauthorblockA{\textit{School of Computing} \\
\textit{University of Georgia}\\
Athens, GA, USA \\
karishma.hegde@uga.edu}
\and
\IEEEauthorblockN{Hemadri Jayalath}
\IEEEauthorblockA{\textit{School of Computing} \\
\textit{University of Georgia}\\
Athens, GA, USA \\
hemadri.jayalath@uga.edu}

}

\maketitle

\begin{abstract}
In a world where technology is increasingly embedded in our everyday experiences, systems that sense and respond to human emotions are elevating digital interaction. At the intersection of artificial intelligence and human-computer interaction, affective computing is emerging with innovative solutions where machines are humanized by enabling them to process and respond to user emotions. This survey paper explores recent research contributions in affective computing applications in the area of emotion recognition, sentiment analysis and personality assignment developed using approaches like large language models (LLMs), multimodal techniques, and personalized AI systems. We analyze the key contributions and innovative methodologies applied by the selected research papers by categorizing them into four domains: AI chatbot applications, multimodal input systems, mental health and therapy applications, and affective computing for safety applications. We then highlight the technological strengths as well as the research gaps and challenges related to these studies. Furthermore, the paper examines the datasets used in each study, highlighting how modality, scale, and diversity impact the development and performance of affective models. Finally, the survey outlines ethical considerations and proposes future directions to develop applications that are more safe, empathetic and practical. 
\end{abstract}


\section{Introduction}
Human emotions are a basic part of our lives, as they influence how we think, behave, connect with others, and interact with the environment. As technology evolves, it increasingly aims to accommodate human emotions and create a better user experience. For instance, modern smart cars are designed to automatically adjust cabin temperature based on external weather conditions, creating a more pleasant environment and positively influencing the mood of passengers \cite{temperature}. Research findings tied to emotions and decision making, coupled with the increasing integration of technology in our daily lives, have led to the further development of fields such as human-computer interaction \cite{hci-rise}. 

As emotions play a critical role in various aspects of life, understanding and managing them effectively has become essential in both personal and professional settings. Human-Computer Interaction (HCI), an interdisciplinary field of cognitive science and engineering focuses on how users perceive systems, and how this interaction could be made more convenient\cite{hci}. In parallel, Artificial Intelligence (AI) is increasingly developing and finding its way into different fields. With the understanding on importance of emotions and the increasing involvement of technology in the daily lives of humans, AI is being used to achieve the goals of HCI through various applications. This effort is moving AI towards interpreting and predicting emotions. The convergence of HCI and AI has given rise to the field of Affective Computing.

This survey gives an idea on the history and development of Affective Computing and focuses on analyzing recent developments in the field of Affective Computing (AC) that are tied to interpreting, predicting and handling human emotions. It has been observed in the study that a vast number of Affective Computing applications being developed to simulate the role of therapists and emotion sensing. By analyzing recent developments in the field, this work seeks to answer how AI-driven technologies can be leveraged to assist users in dealing and enhancing emotional well-being, what challenges remain in their adoption, and what ethical considerations must be addressed to ensure equitable and impactful solutions.

\section{Background}

In D’Mello et al.'s handbook\cite{handbook}, Affective Computing is best described as a multidisciplinary field that integrates insights from engineering, robotics, psychology, neuroscience, and other domains. The introduction of AC has been attributed to Rosalind Picard back in 1995. Initial research around the influence of emotions was done in the field of learning \cite{affective-roaslind}. In their book 'Affective Computing', it is described how learning is often obstructed due to confusion and frustration. They described that the best teachers are observing the pupil's affective state and modify their teaching to work through moments of distress, making the learning process both better and more comfortable. To drive this statement home, they presented the example of a piano tutor system may adjust its teaching speed based on the learner's emotional signals-slowing down and giving support if frustration is detected, or even pushing the learner toward more difficult tasks if it senses engagement. The principles of such tutors did not pertain only to musical education but also to various domains. Some instances mentioned in the book are language learning, gaming, and software training. This instance highlighted the importance of proactive and adaptive responses tailored to the user's affective state, emphasizing that understanding emotions is critical for improving the learning experience.

Another important observation made as part of early Affective Computing research was that the usefulness of measuring emotions. AC owes its origin to the fact that emotions could be quantified using \textbf{biosensors}. A biosensor is a device composed of a biological sensing element that mostly relies on physical contact with the subject to measure electrical signal \cite{biosensor}. The use of biosensors is not new and has been regularly used in the field of medicine and psychology \cite{biosensor-emotion}. Physiological data such as heart rate and skin conductance were found to hold clues regarding emotional states. This discovery led the way in building systems that could evolve on the basis of human emotions.

Emotions can be measured through biosensors by analyzing various physiological signals that correlate with emotional states. Table \ref{biosensors} summarizes some of the popular biosensors and the emotions they focus on. Biosensors have been used to understand and correlate human emotion and biological response. Today, this application is integrated with AI, whose core ability to learn and adapt. This has created a field, AC, which places the emphasis on emotion-centered technology. 

\begin{table}[htbp]
\caption{Biosensors and Emotions}
\begin{center}
\begin{tabularx}{\linewidth}{|X|X|X|}
\hline
\textbf{Physiological Signal} & \textbf{Biosensor} & \textbf{Emotions Measured} \\
\hline
Electroencephalography (EEG) & EEG Sensors & Stress, Relaxation, Excitement \\
\hline
Electrocardiography (ECG) & ECG Sensors & Stress, Anxiety, Calmness \\
\hline
Electromyography (EMG) & EMG Sensors & Joy, Anger, Sadness \\
\hline
Galvanic Skin Response (GSR) & GSR Sensors & Arousal, Stress, Excitement \\
\hline
Photoplethysmography (PPG) & PPG Sensors & Heart Rate Variability linked to Emotional States \\
\hline
Skin Temperature & Thermistors or Infrared Sensors & Stress, Arousal \\
\hline
Respiration Rate & Respiratory Inductive Plethysmography & Calmness, Stress, Excitement \\
\hline
\end{tabularx}
\label{biosensors}
\end{center}
\end{table}

Speaking about the advancements that brought more attention to the field of Affective Computing, in 2010, IEEE launched the flagship journal\textit{IEEE Transactions on Affective Computing} \cite{journal}. It is dedicated to presenting research on systems concerned with human emotions and related affective phenomena.

It is worth mentioning that it was observed in the survey that a significant portion of recent research in Affective Computing focuses on evaluating the ability of AI chatbots, particularly ChatGPT models, to accurately identify the emotions of the user. The studies employ a variety of datasets that are in different formats such as text, audio, video and speech. In the upcoming sections, we describe the key contributions, methodology and outcome of each surveyed work. 

\section{Survey Methodology}
The selection process for this survey focused on identifying recent advancements in Affective Computing (AC) applications. The papers have been selected from renowned journals and research databases including IEEE Xplore, ACM Digital Library, arXiv, and SpringerLink. A considerable number of applications revolve around popular large language models (LLMs) and use deep learning neural networks. During the selection process, it was kept in mind to pick applications that have made unique approaches and impactful contributions, such as usage of niche datasets or developing on state-of-the-art (SOTA) systems.

The research work selected as part of this study mainly focus on the following areas of Affective Computing:
 \begin{itemize}
    \item Chatbots for Emotion Recognition and Sentiment Analysis
    \item Affective Computing in Safety Technologies
    \item Therapy and Mental Health Applications
     \item Multimodal Approaches to Sentiment and Emotional Analysis
 \end{itemize}

Section IV focuses on explaining the goals, technologies, techniques and important results for each paper. Section V highlights the technological strengths of each paper. Next, Section VI gives information about the datasets that were used to train the AI models for various affective computing applications. Next, Section VII describes some of the commonly used metrics and procedures to evaluate Affective Computing systems, as well as effective and innovative methods used in the papers analyzed in this survey. In Section VIII, some of the weaknesses mentioned in each work are highlighted, as well as those identified in this survey. In Section IX some insight on Ethical and Societal Considerations to keep in mind during research in the area of affective computing is given. Finally, the paper concludes with some suggestions for future research in the area.
\section{Applications and Approaches}
This section focuses on the goals and methodologies of selected recent papers in the area of affective computing, categorized based on their primary approach. It highlights the methodology applied to explain what technology and methods have been used. Finally, the key results and contributions of each work is outlined.

\subsection{Chatbots for Emotion Recognition and Sentiment Analysis}
This section focuses on describing research that develop and/or evaluate chatbots for affective computing tasks and emotional analysis. A large chunk of work is based on evaluating different versions of GPT models and the popular chatbot - ChatGPT. Table \ref{chatbot-summary} summarize the goals, methodology and results of each paper in this section.

\begin{table*}[htbp]
\centering
\caption{Summary of Research in the area 'Chatbots for Emotion Recognition and Sentiment Analysis'}
\begin{tabular}{|p{3cm}|p{3cm}|p{5cm}|p{5cm}|}
\hline
\textbf{Paper Title} & \textbf{Goal} & \textbf{Methodology} & \textbf{Results \& Key Findings} \\
\hline
\textit{RVISA: Reasoning and Verification for Implicit Sentiment Analysis\cite{RVISA}} & Implicit Task Analysis (ISA) - recognize hidden sentiment using a structured reasoning and verification process. & Combines the strengths of DO LLMs (for generating reasoning) and ED LLMs (for learning from explanations) for implicit task analysis, achieving SOTA performance. & RVISA had the best accuracy and Macro-F1 score against 7 models. \\
\hline
\textit{GPTEval: A Survey on Assessments of ChatGPT and GPT-4 \cite{GPTEval}} &  Evaluates GPT-3.5 on general NLP tasks and GPT-4 for advanced reasoning & Prompt engineering (zero-shot \& few-shot), comparison with RoBERTa\cite{roberta} and human evaluation & ChatGPT and GPT-4 is strong in language understanding and scientific knowledge, weak in fact verification \& multi-step reasoning. \\
\hline
\textit{Is ChatGPT a Good Sentiment Analyzer? \cite{sentiment-analyzer}} & Tests ChatGPT’s ability in sentiment analysis tasks in text-based inputs. & Given 7 sentiment tasks in zero-shot prompting format with results compared with fine-tuned BERT\cite{BERT} & Performs well in zero-shot settings, but fine-tuned models outperform ChatGPT in structured tasks. \\
\hline
\textit{Is ChatGPT a Good Personality Recognizer? \cite{personality-recognizer}} & Tests ChatGPT’s ability to classify personality traits & Three prompting strategies (zero-shot, zero-shot CoT, one-shot), compared to RNNs \& RoBERTa & ChatGPT is strong in reasoning but shows demographic bias. \\
\hline
\textit{Is ChatGPT a General-Purpose NLP Task Solver? \cite{nlp-solver}} & Evaluates ChatGPT on various NLP tasks & Prompting strategies like CoT, zero-shot prompting & Strong in reasoning tasks but not always superior to fine-tuned models. \\
\hline
\textit{A Wide Evaluation of ChatGPT on Affective Computing Tasks \cite{gpt-wide-eval}} & Evaluates ChatGPT for 13 affective computing tasks & Fixed prompt format for classification, regression modeled as ranking tasks &  Performs well in sentiment \& toxicity detection but weak in implicit emotional cues. \\
\hline
\textit{Sentiment Analysis in the Era of LLMs: A Reality Check \cite{sentiment-llm}} & Compares LLMs \& SLMs in sentiment tasks & Evaluates 7 sentiment task categories using 26 datasets & LLMs excel in zero-shot learning but underperform in structured sentiment tasks. \\
\hline
\textit{Will Affective Computing Emerge from Foundation Models? \cite{first-gpt}} & Evaluates ChatGPT on affective computing tasks like personality \& suicide detection & Comparison with RoBERTa, Word2Vec, BoW & ChatGPT performs decently but lags behind fine-tuned models. \\
\hline
\end{tabular}
\label{chatbot-summary}
\end{table*}

As per OpenAI \cite{openai}, by analyzing vocal tones, facial expressions, and linguistic patterns, GPT-4o can interpret user emotions and generate contextually appropriate responses. This multimodal integration enhances the naturalness and empathy of human-computer interactions. GPT models, such as GPT-3.5, GPT-4, and GPT-4o, are built on Transformer architectures that is based on the self-attention mechanism. It allows the model to asses the importance of different words and sequences in the sentence input.\cite{transformers} This enables the model to capture sentiment dependencies (e.g., sarcasm, negation, or contextual shifts). Additionally, GPT models are trained on vast sentiment-labeled data to learn context-dependent emotional cues.

W. Lai et al.'s work develops \textbf{implicit sentiment analysis (ISA)} system named \textit{RVISA (Reasoning and Verification for Implicit Sentiment Analysis)} \cite{RVISA}. They propose a new approach to ISA by developing two-stage reasoning and verification framework to recognize hidden sentiment. The approach involves reinforcing an \textbf{encoder-decoder (ED)} LLM model with the natural language generation capabilities of a \textbf{decoder-only (DO)} LLM model. 

While DO models (e.g. GPT-3.5-turbo and Vicuna-13B) are excellent at generating text and justifying the outputs with the help of prompts, they may not always issue correct outputs. Hence, the authors run the DO model through a \textit{Three-Hop Reasoning (TH-RE)} prompting mechanism. In the TH-RE method, the model is prompted to identify the aspect of the sentiment first, then determine the opinion related to the aspect, and finally infer the sentiment polarity. This allows the model to generate conclusions supported by reasoning making it more reliable. Next, the Flan-T5 ED model, which is skilled at learning with labeled data is trained with the reasoning that was generated by the DO through TH-RE, thereby making the training process based on context-based data. This helps the model better infer implicit sentiments without relying on explicit cue words. After this, the paper describes running the fine-tuned ED through a simple answer-based system to verify the reasoning generated by the model for its outputs. This process acts as a filtering process to ensure that the ED model learns only from correct reasoning patterns, thereby further increasing the accuracy of implicit sentiment analysis performance.

The model was evaluated on 2 benchmark datasets, namely - \textbf{Restaurant Dataset (SemEval-2014) and Laptop Dataset (SemEval-2014)}. Results were compared against seven state-of-the-art baseline models, including BERT-based approaches and THOR (a Chain-of-Thought model). The performance is measured using accuracy and macro-F1 Score. RVISA achieved the best performance in both measurements. Another observation, where the verification step was removed, resulted in a drop in the Macro-F1 scores. This confirmed its importance in improving the output quality.

R. Mao et al. presented in their work \textit{'GPTEval: A survey on assessments of ChatGPT and GPT-4'} the methods and results from previous evaluations of the GPT model\cite{GPTEval}. The paper reviews GPT-3.5 on general \textbf{Natural Language Processing (NLP) tasks} and ethical compliance in terms of facts and accuracy of presented information. GPT-4 is benchmarked on \textbf{advanced reasoning} focusing on areas of language proficiency, reasoning abilities, scientific knowledge, ethical considerations and multi-turn dialogues. Techniques evaluated included:
\begin{itemize}
    \item  Prompt engineering (zero-shot and few-shot setups).
    \item Comparisons with fine-tuned models like Robustly Optimized BERT Pretraining Approach (RoBERTa).\footnote{ RoBERTa is an improved version of BERT (Bidirectional Encoder Representations from Transformers) developed by Meta AI (Meta AI). RoBERTa enhances BERT by using methods such as removing the next sentence prediction (NSP) task during training and using a larger batch size.}
    \item Human evaluations for tasks involving creativity and ethical considerations.
    \item External tools integrated for specific benchmarks (e.g., OpenAI's API for dynamic queries).
\end{itemize}

The study revealed that ChatGPT and GPT-4 are robust in language understanding and generation. However, they underperformed in domain-specific areas compared to fine-tuned models. ChatGPT showed promising results in giving responses on  scientific knowledge and theoretical developments, but underperformed multi-step reasoning tasks and the verification of facts. In this regard, the author pointed out ethical issues.

In the work of Z. Wang et al., \textit{'Is ChatGPT a Good Sentiment Analyzer? A Preliminary Study'} \cite{sentiment-analyzer}, the authors work on a prefatory study to test the capability of ChatGPT in identifying the \textbf{context and emotions} in the given text. In the paper, seven typical sentiment analysis tasks were conducted. The seven tasks are:

\begin{itemize}
    \item Sentiment Classification (SC)
    \item Aspect-Based Sentiment Classification (ABSC)
    \item End-to-End Aspect-Based Sentiment Analysis (E2E-ABSA)
    \item Comparative Sentences Identification (CSI)
    \item Comparative Element Extraction (CEE)
    \item Emotion Cause Extraction (ECE)
    \item Emotion-Cause Pair Extraction (ECPE)  
\end{itemize}

Using 17 benchmark datasets, GPT-3.5-turbo-0301 was executed using the in a complete \textbf{zero-shot method} with advanced prompting techniques. The authors have changed the meaning of the same prompts toward specific tasks involved in extracting sentiment, identifying comparative elements. The performance of the model as compared to that of fine-tuned BERT, BERT-base-uncased. The fine-tuned models outperformed ChatGPT as they were fine-tuned to handle the specific type of task. However, authors state that ChatGPT demonstrated strong zero-shot abilities in sentiment classification, CSI, and ECE. The study highlights ChatGPT’s robustness in polarity shift detection and its generalization ability across open-domain sentiment analysis, despite its limitations in structured sentiment extraction.

Y. Ji et al. in \textit{'Is ChatGPT a Good Personality Recognizer? A Preliminary Study'} study the performance of ChatGPT in handling \textbf{text-based personality recognition tasks} \cite{personality-recognizer}. The model is compared to traditional neural networks (NN) in the form of Residual Neural Networks (RNNs) with GloVe embeddings, fine-tuned transformer-based RoBERTa and SOTA model. Using 2 personality recognition datasets (Essay and PAN) and \textbf{3 prompting strategies (zero-shot prompting, zero-shot Chain-of-Thought (CoT) prompting, and one-shot prompting)} authors checked effectiveness on 2 downstream tasks - a \textbf{sentiment classification} task and a \textbf{stress prediction} task. Results showed that ChatGPT, especially when \textbf{zero-shot CoT prompting}, is considerably accurate at personality recognition tasks. It also showed accuracy in providing explanations highlighting its capability in natural language generation. Authors also stated that the model showed bias in predictions related to sensitive demographic attributes such as gender and age. Further, utilizing ChatGPT's personality recognition capability enhanced its performance on related downstream tasks such as sentiment classification and stress prediction.

\textit{'Is ChatGPT a General-Purpose Natural Language Processing Task Solver?'}, by C. Qin et al.'s contributes an evaluation study of ChatGPT on natural language processing tasks \cite{nlp-solver}. Using prompt engineering like \textbf{zero-shot learning and CoT prompting}, ChatGPT was evaluated across 20 popular NLP datasets covering 7 representative task categories. The 7 tasks are:
\begin{itemize}
    \item Reasoning
    \begin{itemize}
        \item Arithmetic Reasoning
        \item Commonsense Reasoning
        \item Symbolic Reasoning
        \item Logical Reasoning
    \end{itemize}
    \item Natural Language Inference (NLI)
    \item Question Answering (QA) / Reading Comprehension
    \item Dialogue
    \item Summarization
    \item Named Entity Recognition (NER)
    \item Sentiment Analysis
\end{itemize}
Performance comparisons were conducted with the earlier versions of GPT, including GPT-3.5-text-davinci-003, and fine-tuned versions like PaLM and T0. The results showed that ChatGPT outperformed GPT-3.5 in arithmetic and logical reasoning, qualitatively excellent in multi-step problem-solving. It also performed impressively on NLI, QA, and dialogue reasoning, highlighting its  ability to understand and produce complex responses. It also displayed enhanced capability in self-correction of errors compared to GPT-3.5. 

In the paper titled \textit{'A Wide Evaluation of ChatGPT on Affective Computing Tasks'}, M. Amin et al. presented a broad assessment of ChatGPT models, notably GPT-3.5 and GPT-4, with respect to 13 affective computing tasks\cite{gpt-wide-eval}. The tasks covered included:
\begin{itemize}
    \item Aspect Extraction
    \item Aspect Polarity Classification
    \item Opinion Extraction
    \item Sentiment Analysis
    \item Sentiment Intensity Ranking
    \item Emotion Intensity Ranking
    \item Suicide Tendency Detection
    \item Toxicity Detection
    \item Well-being Assessment
    \item Engagement Measurement
    \item Personality Assessment
    \item Sarcasm Detection
    \item Subjectivity Detection
\end{itemize}
The prompt followed a fixed format which first instructed ChatGPT its role, given the problem description. It then instructed that given the user input, the expected output is a label. Then, the dos and don'ts of the output format was defined. Next, the authors introduced a framework for testing ChatGPT on \textbf{regression-based problems} as querying an exact score is difficult due to scale variations and subjective dataset annotations. The framework formatted the regression problem as a \textbf{pairwise ranking classification task}. The model had to determine whether one label is greater than another. 

ChatGPT was competed against traditional NLP methods, including end-to-end recurrent neural networks and transformers. The results showed that GPT-3.5 and especially GPT-4 performed very well in tasks related to sentiment, emotions, and toxicity. However, the model underperformed with implicit signal tasks like the measurement of engagement and subjectivity detection. This study suggests that while ChatGPT is promising in affective computing, there is definitely room for improvement with regard to the handling of tasks that require the understanding of implicit emotional cues. 

In the paper \textit{'Sentiment Analysis in the Era of Large Language Models: A Reality Check'}, W. Zhang et al. deeply studies the performance of different language models in a wide variety of tasks around sentiment analysis \cite{sentiment-llm}. The different tasks that were considered for this study included:
\begin{itemize}
    \item Sentiment Classification at a document-level, sentence-level and aspect-level.
    \item Aspect-based Sentiment Analysis (ABSA) (fine-grained sentiment analysis)
    \item Multifaceted Analysis of Text
    \begin{itemize}
        \item Implicit Sentiment Analysis 
        \item Hate Speech Detection
        \item rony Detection
        \item Offensive Language Identification
        \item Stance Detection 
        \item Comparative Opinion Mining
        \item Emotion Recognition 
    \end{itemize}
\end{itemize}
The models evaluated included both \textbf{large language models (LLMs)} such as and \textbf{small language models (SLMs)}. LLM models considered were the \textbf{Flan-T5 (XXL, 13B) and Flan-UL2 (20B)} from the Flan model family and OpenAI models \textbf{ChatGPT (gpt-3.5-turbo3)} and \textbf{text-davinci-003 (175B, GPT-3.5 family)}. For SLMS, the fine-tuned \textbf{T5 (Large, 770M)} trained using the Adam optimizer. For the evaluation process, 26 datasets in conventional sentiment classification, aspect-based sentiment analysis was employed. 

The results showed that LLMs perform well on simpler tasks (e.g., binary sentiment classification), but fall short in more complex tasks requiring deeper understanding or structured sentiment information (e.g. fine-grained sentiment analysis and ABSA). The \textbf{zero-shot learning} settings favored LLMs but did not prove the same for SLMs, particularly in structured sentiment tasks. Similarly, in the \textbf{few-shot learning} settings, LLMs outperform SLMs, showcasing their ability to generalize from limited labeled data. However, SLMs outperform  LLMs with the same few-shot learning when fine-tuned on large-scale domain-specific data, especially in structured sentiment tasks like ABSA. The authors discussed limitations in current model evaluation practice as they are not structured and standardized to test the model's capabilities in full depth. Hence, the authors introduced the \textbf{\textsc{SentiEval}} benchmark. The goal of this benchmark is to mitigate issues related to inconsistent prompt engineering and biased evaluation methods in past studies.

Another work, which evaluated ChatGPT's text classification capabilities is M. Amin et al.'s \textit{'Will Affective Computing Emerge from Foundation Models and General AI? A First Evaluation on ChatGPT'} \cite{first-gpt}. This paper focused on three affective computing tasks: \textbf{Big-Five personality prediction, sentiment analysis, and suicide tendency detection}. Comparisons have been made among three models. Firstly, a transformer architecture based \textbf{RoBERTa} model, pre-trained on various books, the Wikipedia knowledge base, english news data, Reddit posts, and stories is considered. Secondly, the Word2Vec model, which is pre-trained on a large dataset derived from Google news is considered. The final model is a basic BoW model which only uses internal data for training. The evaluation metrics include classification accuracy and Unweighted Average Recall (UAR), with randomized permutation tests used for statistical significance. The results showed that the RoBERTa model fine-tuned on specific downstream tasks generally performed better than the rest, while ChatGPT showed decent results close to the baselines Word2Vec and BoW. Again, ChatGPT was able to be robust against noisy data where the performance of the Word2Vec model went down. The authors claim that ChatGPT is a good generalist model that achieves reasonable performance on many tasks without special training but still far from the top models which are fine-tuned on a particular task. 

\subsection{Multimodal Approaches to Sentiment and Emotional Analysis}
This section covers studies that extend beyond the purely textual sentiment analysis into including other modalities like contextual knowledge and visual cueing. Unlike some of the previous works discussed, which are based on textual cues only, multimodal techniques consider visual expressions, acoustic features, and contextual knowledge for applications in areas like emotion detection, sarcasm identification, and sentiment classification. The table \ref{multimodal-summary} summarize the goals, methodology and results of each paper in this section.

\begin{table*}[htbp]
\centering
\caption{Summary of Research in the area 'Multimodal Approaches to Sentiment and Emotional Analysis'}
\begin{tabular}{|p{4cm}|p{3cm}|p{5cm}|p{4cm}|}
\hline
\textbf{Paper Title} & \textbf{Goal} & \textbf{Methodology} & \textbf{Results \& Key Findings} \\
\hline
\textit{Novel Speech-Based Emotion Climate Recognition in Peers’ Conversations Incorporating Affect Dynamics and Temporal Convolutional Neural Networks \cite{peers}} & Speech-based deep learning AI system that observes group conversations to assess the overall emotional climate. & Extracts features from speech data using MFCCs\cite{MFCC} and applied to a TCNNs which predicts emotions. & SOTA accuracy of 83.3\% for arousal and 80.2\% for valence.\\
\hline
\textit{VyaktitvaNirdharan: Multimodal Assessment of Personality and Trait Emotional Intelligence \cite{VyaktitvaNirdharan}} & Design and Prototype of a multi-model and multitask assessment of Personality and Trait EI in the language Hindi. & The Vyaktitva dataset is created by recording participants' conversations, and a self-assessed questionnaire for Big Five Personality \cite{bigfiveLiverpool} and EI Trait, used as training labels. The dataset is processed to extract visual, acoustic, and linguistic features. The extracted multimodal features were used to train deep learning models for personality and EI prediction. & The deep learning pipeline effectively predicted Big-Five personality traits with high accuracy. \\
\hline
\textit{KnowleNet: Knowledge fusion network for multimodal sarcasm detection \cite{knowleNet}} & Model that incorporates prior knowledge and cross-modal semantic similarity detection to better detect sarcasm in social media data. & Uses ConceptNet\cite{conceptnet} for prior knowledge and contrastive learning to increase the difference between sarcastic and non-sarcastic samples. & Attained SOTA performance on publicly available benchmark datasets at 88.87\%\\
\hline
\textit{Sarcasm Detection in News Headlines using Supervised Learning \cite{sarcasm-news}} & Classify news headlines into either sarcastic or not sarcastic by leveraging seven machine learning algorithms & Models were trained and tested on the labeled data regarding the evaluation of their abilities to find sarcasm. This data was used with the feature representations, BoW and context-independent-fastText embeddings, and context-dependent, BERT embeddings. & RoBERTa model performed best with a Micro-F1 score of 93.11\% on the test data, signifying the importance of context-dependent features for sarcasm detection.\\
\hline
\textit{The Biases of Pre-Trained Language Models: An Empirical Study on Prompt-Based Sentiment Analysis and Emotion Detection \cite{biases}} & Throws light on presence of biases in prompt-based classification in pre-trained language models. & Variants of BERT, RoBERTa, ALBERT and BART were tested with varying number of label classes and choices of emotional label words for performance and fairness. Prompts consisted of zero-shot classification. & Results confirmed the hypotheses of presence of biases in results with varying prompts, raising concerns for standardized and ethical prompt-based classification of PLM models. \\
\hline
\textit{SenticNet 7: A commonsense-based neurosymbolic AI framework for explainable sentiment analysis \cite{SenticNet}} & Commonsense-based neurosymbolic AI framework to overcome issues in AI models & Combine a Subsymbolic AI to learn patterns from text and a Symbolic AI to extract context. & Proved to handle sentiment analysis in a trustworthy and transparent way and also accurately identify multi-word expressions containing polarity disambiguation.\\
\hline
\end{tabular}
\label{multimodal-summary}
\end{table*}

G. Alhussein et. al.'s work \textit{'Novel Speech-Based Emotion Climate Recognition in Peers’ Conversations Incorporating Affect Dynamics and Temporal Convolutional Neural Networks'} focuses on a unique problem statement primarily using data in the form of speech as the input for the model \cite{peers}. The paper presents a \textbf{speech-based AI system} that observes \textbf{group conversations} to assess the overall emotional climate of the conversation. Using deep learning and speech processing, the model outperforms previous methods and has applications in therapy, negotiations, and emotion-aware AI. The proposed system first begins with a process known as \textbf{Mel-Frequency Cepstral Coefficients (MFCCs)} to extract features from speech data while filtering the noise. Next, it is applied to a neural network. The structure of the ML model used is \textbf{Temporal Convolutional Neural Networks (TCNNs)}. to model affect dynamics. Next, to capture the variation in the emotions during the conversation, TCNN analyzes dynamic emotional patterns from speech features. This system is tested across three benchmark datasets, K-EmoCon (multimodal emotional data including physiological signals), IEMOCAP (actor-based emotional speech dataset) and SEWA (cross-cultural dataset with speech and video). The model is evaluated to check it's accuracy for predicting the arousal and valence levels. The model showed promising state-of-the-art results, with 83.3\% accuracy for arousal and 80.2\% for valence.

The paper \textit{VyaktitvaNirdharan: Multimodal Assessment of Personality and Trait Emotional Intelligence} \cite{VyaktitvaNirdharan} by M. Leekha et al proposes a \textbf{multi-modal-multitask learning (MM-MTL) system} for predicting \textbf{Big-Five personality} and \textbf{Emotional Intelligence (EI)} traits using the Hindi language in near real-time. The authors highlight the use of personality assessment in better understanding of personnel in fields like military, National Aeronautics and Space Administration (NASA), corporate organizations, educational institutions and psychology. It is also used in targeting products, personalizing experiences and services. The authors emphasize that personality and Trait EI are correlated. In this regard, they have developed a multitask model that simultaneously assigns the personality as well as predicts the EI trait of a user. The developed model is a first of its kind because not only is the EI predicted during real-time multimodal data, it is being done in a lesser represented language i.e. Hindi. 

The study begins with the data collection process for the Vyaktitva dataset, which records a 1:1 conversation between selected participants. Next, these participants complete a standardized questionnaire for Big Five Personality and EI Trait. As the personality and EI assessment did not involve an external rater, it eliminated the labeling bias. The dataset was processed to extract visual, acoustic, and linguistic features. The extracted multimodal features were used to train deep learning models for personality and EI prediction. The \textbf{VGG-16} (Pre-trained CNN Model) was employed for processing key frame video data. The \textbf{LibROSA} library was used for processing speech features. The deep learning pipeline effectively predicted Big-Five personality traits with high accuracy. The authors state that the model performed better in predicting certain personality traits that were characterized to be more explicit. The prototype was also tested with users from two stakeholders in the recruiting and psychology field. It was well received and demonstrated practical potential in the two domains.

In the work \textit{'KnowleNet: Knowledge fusion network for multimodal sarcasm detection'}, authors Yue T. et al. propose a model that incorporates prior knowledge and cross-modal semantic similarity detection to better detect sarcasm in social media data \cite{knowleNet}. The paper addresses two challenges in the detection of sarcasm: 
\begin{itemize}
    \item Enhancing the state-of-the-art methods by leveraging commonsense knowledge via \textbf{ConceptNet} knowledge base.
    \item Addressing inconsistencies between image and text modalities.
\end{itemize}
It also employs contrastive learning to increase the difference between sarcastic and non-sarcastic samples. The performance evaluations conducted on publicly available benchmark datasets proved that KnowleNet attained SOTA performance at 88.87\%, highlighting the effectiveness of prior knowledge combined with multimodal data in performing sarcasm detection. In fact, the ablation study in the paper showed that without word-level or sample-level semantic similarity detection modules, plus the triplet loss, performance significantly reduces; hence, all are indispensable.

In the paper, \textit{'Sarcasm Detection in News Headlines using Supervised Learning}', the authors attempted to classify news headlines into either sarcastic or not sarcastic by leveraging seven machine learning algorithms, and their corresponding different types of features to identify sarcasm \cite{sarcasm-news}. The models that were evaluated are:
\begin{itemize}
    \item Naïve Bayes-support vector machine
    \item Logistic Regression
    \item Bidirectional Gated Recurrent Units,
    \item Bidirectional encoders representation from Transformers (BERT)
    \item DistilBERT
    \item RoBERTa
\end{itemize}
The dataset used contained both sarcastic (source: The Onion) and non-sarcastic (source: HuffPost) news headlines. Preprocessing steps like Unicode correction, token expansion, and punctuation removal were done to prepare the data. Models were trained and tested on the labeled data regarding the evaluation of their abilities to find sarcasm. This data was used with the feature representations, BoW and context-independent-fastText embeddings, and context-dependent, BERT embeddings. Their experimental results demonstrated that the RoBERTa model, which captures the bidirectional semantic context, outperformed the other models with a Micro-F1 score of 93.11\% on the test data. This suggests that context-dependent features are more apt for the purpose of sarcasm detection in news headlines.

It is noteworthy that the RoBERTa model has generally better than most other models across multiple research papers in sentiment analysis and emotion detection tasks.

R. Mao et al. presented a study in the paper \textit{'The Biases of Pre-Trained Language Models: An Empirical Study on Prompt-Based Sentiment Analysis and Emotion Detection'} focusing on \textbf{pre-trained language models} \cite{biases}. The rising availability of pre-trained models have allowed sentiment analysis and emotion detection through prompt based inputs. This research focuses on providing comprehensive insight into the nature of the different biases present in prompt-based classification, highlighting many research challenges on this topic. The models investigated in this work are \textbf{BERT, RoBERTa, ALBERT, BART}. The smaller and larger variants of such models were considered. The methodology of the experiment focused on investigating how factors such as the number of label classes and choices of emotional label words affect performance and fairness for such models. Tasks utilized prompts with templates for zero-shot classification- (e.g. \textit{I feel [MASK]}). Results reveled that the PLMs do have biases regarding the above factors. That means the sentiment and emotion classification by these models is not guaranteed to be accurate and unbiased. The authors called for awareness of these biases, which must be mitigated to ensure that ethical deployment of PLMs occurs in affective computing applications.

C. Erik et al. in the paper \textit{'SenticNet 7: A Commonsense-based Neurosymbolic AI Framework for Explainable Sentiment Analysis'} \cite{SenticNet} have proposed a \textbf{commonsense-based neurosymbolic AI framework}. It aims to overcome identified issues like trustworthiness, interpretability, and explainability in AI models in the context of sentiment analysis. A neurosymbolic framework is a combination of \textbf{neural networks (subsymbolic AI)} and \textbf{symbolic AI}. The SenticNet 7 framework integrates:
\begin{itemize}
    \item \textbf{Subsymbolic AI} (Auto-regressive language models, kernel methods): This helps the model Learn patterns from text, generating an unsupervised and reproducible model.
    \item  \textbf{Symbolic AI} (Commonsense knowledge graphs): This enables the model to extract structured meaning and relationships from the text.
\end{itemize}
SenticNet 7 was tested against 20 popular English lexica in sentiment analysis on 10 different datasets for proving its usefulness. The authors then showed that their method overcame the identified limitations in a manner appropriate for conducting sentiment analysis in a trustworthy and transparent way. It was even able to handle multi-word expressions containing polarity disambiguation (e.g. dead vs dead-right).

\subsection{Applications of Affective Computing in Therapy and Mental Health}

This section summarizes research on Affective Computing for mental health support, including AI-based therapy and counseling systems. These will be designed not only to interpret the emotions correctly but also to provide personalized emotional support, well-being interventions, and therapy-oriented interactions, while they are required to make users who use these systems feel that they get a human-like experience. The table \ref{therapy-summary} summarize the goals, methodology and results of each paper in this section.

\begin{table*}[htbp]
\centering
\caption{Summary of Research in the area 'Applications of Affective Computing in Therapy and Mental Health'}
\begin{tabular}{|p{4cm}|p{3cm}|p{5cm}|p{4cm}|}
\hline
\textbf{Paper Title} & \textbf{Goal} & \textbf{Methodology} & \textbf{Results \& Key Findings} \\
\hline
\textit{Generative Ghosts: Anticipating Benefits and Risks of AI Afterlives} \cite{generativeghosts} & To conceptualize and define generative ghosts and explore their design space, potential benefits, and associated risks in order to guide future research and ethical development in AI-driven posthumous representation. & Conceptual analysis and design space exploration based on literature review, case studies, and ethical reflection to define generative ghosts and assess their benefits and risks. & Outlines a design framework with dimensions like provenance, embodiment, and evolution, and highlights potential benefits and critical risks.\\
\hline
\textit{The "Conversation" about Loss: Understanding How Chatbot Technology was Used in Supporting People in Grief \cite{grief-chatbot}} & Investigates the effectiveness of usage of chatbots in comforting grieving users. & Participants used two chatbots followed by a semi-structure interview to share their thoughts on the experience. & Participants indicated a positive impact, though emphasizing the role of real-life interactions, which is irreplaceable. \\
\hline
\textit{EMMA: An Emotion-Aware Wellbeing Chatbot \cite{emma}} & Develop EMMA, an emotionally conscious chatbot for mental health issues. & Two-week study. Week 1 involved collecting data by having participants interact with EMMA. In week 2, the data was fed into an ML model, and based on the inferred mood of the user, it suggested mindful activities to alleviate the user's emotional state. & The chatbot was reported to be positively effective by users. The authors also suggested guidelines for designing emotion-aware bots.\\
\hline
\textit{An Affectively Aware Virtual Therapist for Depression Counseling \cite{virtual-therapist}} & Design an affectively aware system that plays the role of a virtual therapist responding to users' emotional states during therapy sessions. & A virtual agent "ECA", powered by BEAT to simulate a real-life therapist was prototyped to collect input data from users through speech and facial expressions. It would then generate empathetic responses based on the valance and arousal measured. The prototype was tested on a small group of mild to moderately depressed users. & Users reported a minor, but positive impact of the virtual agent that accurately classified the user emotion. \\
\hline
\end{tabular}
\label{therapy-summary}
\end{table*}

AI mental health applications leverage emotion-aware NLP, multimodal affect detection, and user-adaptive responses to improve the effectiveness of engagements. Such systems aim at non-intrusive, accessible, and scalable mental health support by embedding the principles of affective computing to address a wide range of mental health challenges related to emotional distress, grief, and psychological well-being.

The following studies will discuss technology playing the role of therapists by describing their methodologies and results.

The paper \textit{Generative Ghosts: Anticipating Benefits and Risks of AI Afterlives}\cite{generativeghosts} discusses a range of advanced technologies used to create emotionally \textbf{responsive posthumous AI agents}, referred to as generative ghosts. These include large LLMs like GPT-4, PaLM 2, and LLaMA 2, which are capable of mimicking human conversations. Additionally, the paper references multimodal generative models that produce images, avatars and videos (e.g., DALL·E, Make-A-Video, AudioLM), enabling more \textbf{immersive} and \textbf{lifelike interactions}. Tools like OpenAI’s GPTs provide no-code platforms for building such agents, while commercial services like \textit{HereAfter} and \textit{Re;memory} offer end-of-life digital legacy creation. The paper also presents experimental cases such as \textit{Fredbot} and \textit{Roman} bot—chatbots that use personal data of the deceased to generate avatars. These technologies focus on offering emotional support, memorialization, and therapeutic interactions.

In the paper \textit{'The 'Conversation' about Loss: Understanding How Chatbot Technology was Used in Supporting People in Grief'}, A. Xygkou et al. researched how people coping with grief used chatbots \cite{grief-chatbot}. The paper deals with usage of chatbots by users in grief. More specifically, they show that a chatbot could support mourners to cope with their loss by continuing a bond, keeping them company, or offering emotional support. Two platforms were used for this study, which are: 
\begin{itemize}
    \item \textbf{Replika}: A mobile application chatbot that can be customized to play the role of various types of relationships such as a friend, romantic partner or family member. 
    \item \textbf{Project December-Simulation Matrix}: Developed by OpenAI, it combines the capabilities of GPT-2 and GPT-3 to simulate deceased people and allow users to converse with them for solace.
\end{itemize}
 For this study, 10 grieving participants were asked to use both  platforms, which served as training data. The authors investigated how chatbots affect the grieving process with respect to loneliness, closure, identity reconstruction, and social connectedness. Lastly, they explore the advantages and disadvantages of using chatbots as a supplemental intervention tool for therapy and emotional support in grief. Though the benefits were therapeutic in nature, participants mentioned that a combination of using a chatbot with real-life human interactions and professional therapy is crucial. The authors conclude with mentioning potential risks with regard to ethics and a need to study long-term impacts of these systems on the users.

Ghandeharioun et al. presented an emotionally conscious personal assistant in \textit{'EMMA: An Emotion-Aware Wellbeing Chatbot'} \cite{emma}. The work aims at providing mental health interventions through accessible devices such as smartphones. The methodology involves a two-week study with 39 participants to assess the effectiveness of EMMA. Participants interacted with EMMA five times a day in the first week in order to log their emotional states. In parallel, continuous collection of sensor data, which included geolocation and activity level, was applied to gather contextual information about the participants' environmental and behavioral state. In the second week, an ML model was used to infer the mood of participants from the data gathered in the first week, and for the inferred moods, emotionally appropriated micro-activities like mindfulness exercises, positive affirmations, and breathing techniques were provided through this system. These were empathetic and supportive interventions to help people improve their overall well-being. This study showed that EMMA was able to detect the users' mood and provide them with personalized micro-activities to support their mental health.

The paper \textit{'An Affectively Aware Virtual Therapist for Depression Counseling'} by L. Ring et al. describes the design and prototype of an affectively aware system that plays the role of a \textbf{virtual therapist} \cite{virtual-therapist} capable of detecting and responding to users' emotional states during therapy sessions. It is targeted to enhance the limited capabilities of most computer-based traditional \textbf{cognitive behavioral therapy (CBT)} systems that fail to detect a user's emotional state during the therapy sessions. They integrate emotion recognition with CBT to make automated counseling more effective. The prototype designed is based on a manualized CBT intervention developed by a clinical expert. 

A virtual agent called \textbf{Embodied Conversational Agent (ECA)} serves as the therapist, engaging users in structured therapy dialogues. Using the \textbf{Behavior Expression Animation Toolkit (BEAT) }engine, the agent simulates human expressions and body language by translating text to speech and also generates nonverbal behaviors gestures, facial expressions, and posture shifts. Users interact with the agent through speech or a touch-screen mode. The system aims to classify user data three valence categories: happy, neutral, or sad. The classifier was trained using \textbf{OpenSmile} and \textbf{LibSVM} on the Emotional Prosody Speech and Transcripts database. When speech-based affect detection is available, the \textbf{Affdex SDK} is used to observe facial expressions to classify emotions. Responses by the agent were generated based on the valence and arousal levels of user emotions.

The prototype was put through a test with individuals reporting mild to moderate depression. After interacting with the virtual agent, a semi-structured interview was conducted for feedback. Results revealed that the virtual agent did accurately identify and empathize with the users. Although the extent and intensity of change in their mood was minimal, the interaction with the agent left them with a positive impact. Authors concluded the paper with future work involving more extensive studies and enhance the existing system potentially with capabilities of NLP.

\subsection{Affective Computing in Safety Technologies}
This section covers the application of Affective Computing in safety-critical environments where real-time emotion recognition can be used to improve user well-being and decision-making. Among the research areas, one of the most prominent is stress detection while driving since driving performance and road safety are seriously affected by the level of stress and cognitive overload. The table \ref{safety-summary} summarize the goals, methodology and results of each paper in this section.

\begin{table*}[htbp]
\centering
\caption{Summary of Research in the area 'Affective Computing in Safety Technologies'}
\begin{tabular}{|p{4cm}|p{3cm}|p{5cm}|p{4cm}|}
\hline
\textbf{Paper Title} & \textbf{Goal} & \textbf{Methodology} & \textbf{Results \& Key Findings} \\
\hline
\textit{EmoTake: Exploring Drivers’ Emotion for Takeover Behavior Prediction \cite{emotake}} & Proposes EmoTake, a deep learning based system that explores drivers’ emotions to predict takeover readiness, takeover reaction time, and takeover quality & Participants were subjected to a driving simulator, where they watched a video as the NDRT. The facial experssions, head and eye movement, and body posture was captured by a camera which was fed to an iTransformer network for time series forcasting. & Predicts takeover readiness, reaction time and quality with an accuracy of 91.74\%, 88.55\%, and 81.71\% respectively.\\
\hline
\textit{Predicting Driver Self-Reported Stress by Analyzing the Road Scene \cite{driver-stress}} & Determine if visual information about the road scene can be used to determine the driver's emotional state and level of stress. & Three Computer Vision methods were designed and evaluated to asses driver stress: Object Presence Features, End-to-End Image Classification and End-to-End Video Classification & All three approaches demonstrated promising results paving the way for further research and development in the area of driver assistance systems and road safety systems.\\
\hline
\end{tabular}
\label{safety-summary}
\end{table*}

Y. Gu. et al. in \textit{'EmoTake: Exploring Drivers’ Emotion for Takeover Behavior Prediction'} \cite{emotake} work on developing a deep neural network system called EmoTake that records and predicts the drivers’ emotions to measure the takeover readiness, takeover reaction time, and takeover quality. With developments in \textbf{self-driving vehicles}, users are often occupied in non-driving-related tasks (NDRTs), but are required to take over the responsibility of driving based on the road situation. Since the passenger may not be alert during such take-over requests, it may affect the quality of their response, risking safety. Authors state that the nature of the take-over is influenced by the emotional state of the passenger, which is in turn determined by the NDRT they are engaged in. 

This novel study, considers 26 participants in a driving simulator environment developed with the CARLA software. The NDRT for the experiment was to watch selected video clips to evoke different types of emotions. After 4 minutes of the video, the participant was issued a take-over request (TOR). The facial expression, head pose and eye movements, and the body posture was now observed with a camera. This input was fed to an \textbf{iTransformer}, which is a network designed for time-series data forecasting. It would be used to predict the user's valance and arousal levels for assess their readiness to take over driving. A semi-structured interview was conducted after the driving portion of the experiment which inquired the participant's emotions upon watching the videos.

EmoTake achieved high accuracy in predicting driver takeover readiness (91.74\%), reaction time (88.55\%), and takeover quality (81.71\%), and outperformed traditional machine learning methods. The results confirmed that the emotional state of users have a significant impact on takeover behavior. Surprisingly, high-arousal and negative emotions improved response time. Results also implied the effectiveness of multi-channel fusion for this type of safety mechanism in semi-automated vehicles. 

In the paper \textit{'Predicting Driver Self-Reported Stress by Analyzing the Road Scene'}, Cristina Bustos et al. have worked to determine whether visual information from the driving environment could be used to estimate drivers' subjective stress levels \cite{driver-stress}. Using the \textit{AffectiveROAD} dataset, the which was categorized into three discrete classes of stress levels: low, medium, and high the authors outline and assess three approaches in modeling a computer vision to classify driver's level of stress due to a visual scene:
\begin{itemize}
    \item \textbf{Object Presence Features}: This approach involved automatic scene segmentation to identify the presence of objects in the driving environment.
    \item \textbf{End-to-End Image Classification}: In this method, individual frames from the video recordings were used to train a model to classify stress levels directly from static images.
    \item \textbf{End-to-End Video Classification}: This approach extended the image classification method by including temporal information, allowing the model to analyze sequences of frames to capture dynamic aspects of the driving scene.
\end{itemize}

All three approaches demonstrated promising results, suggesting that it is possible to predict the drivers' subjective stress levels using visual information from the driving scene. Assumptions made below were found to correspond to results:
\begin{itemize}
    \item Parking - Low stress
    \item Highway - Medium stress
    \item City Driving - High stress
\end{itemize}

The study concludes that analyzing the visual driving scene can be a valuable method for estimating drivers' stress levels, which could inform the development of advanced driver assistance systems aimed at improving road safety.

\section{Technological Strengths}
This section examines the key technological strengths of the research studies surveyed, highlighting their contributions to sentiment analysis, emotion recognition, and affective computing. In addition to summarizing individual strengths, this section also provides a comparative analysis of different approaches. By evaluating how each method improves upon prior techniques, addresses specific challenges, and outperforms baseline models, this discussion offers insights into its relative advantages.

\subsection{Evaluation and Generalization of GPT Models}
GPTEval: A survey on ChatGPT and GPT-4 assessments by \cite{GPTEval} performed a general review of the GPT model, underlining strengths with respect to general NLP tasks, especially in zero-shot and few-shot learning. This study stressed the importance of having a unified evaluation benchmark in assessing models like GPT more equitably. Similarly, \textit{A Wide Evaluation of ChatGPT on Affective Computing Tasks }\cite{gpt-wide-eval} has demonstrated the capability of GPT-4 in sentiment-related tasks using prompting-based techniques that avoid task-specific training. \textit{Will Affective Computing Emerge from Foundation Models and General AI? }\cite{first-gpt} further justifies the generalization ability of ChatGPT in affective computing, stating its robustness against noisy data. Is ChatGPT a General-Purpose Natural Language Processing Task Solver? by citation nlp-solver, investigates how well the model generalizes across very disparate NLP tasks, placing special emphasis on reasoning, sentiment analysis, and multimodal integration. The model underlines its rich commonsense knowledge that enhances semantic coherence in tasks requiring contextual understanding. It achieves state-of-the-art performance in sarcasm detection: it achieved an accuracy of 88.87\% and an F1-score of 86.33\% on a benchmark dataset. Besides, ChatGPT demonstrates strong integration of textual and visual inputs, outperforming unimodal and competitive multimodal models. The model also generalizes well, yielding 64.35\% on the secondary dataset, which shows the scalability across different data distributions.

The methods used for evaluation are unreliable: significant differences in performances are due to prompt engineering and dataset exposure; these might bias model assessments. They highlight a dire need to develop more transparent and standardized mechanisms of evaluation through which large language models can be suitably reviewed. 

\subsection{Sentiment Analysis and Emotion Detection}

\cite{peers} is a niche application of combining affect dynamics and conversational speech for EC in a deep Neural network. The datasets that are used in this work have addressed the weakness in many existing ones. In particular it is a diverse and multimodal collection that represents cross culture speech data with real test subjects.
Unlike traditional CNN models, the TCNN used considers capturing long-term dependencies in conversations, which is useful when the EC changes over the course of time. The model is also attributed to extract deep hierarchical features that enhance emotion prediction.
It is also equipped to handle conversational data across different languages by generalizing emotional cues in speech.

\textit{Is ChatGPT a Good Sentiment Analyzer? A Preliminary Study} \cite{sentiment-analyzer} establishes ChatGPT’s proficiency in zero-shot sentiment classification, surpassing fine-tuned BERT in handling polarity shifts. Meanwhile, \textit{Sentiment Analysis in the Era of Large Language Models: A Reality Check} \cite{sentiment-llm} introduces SENTIEVAL, a unified benchmark for sentiment evaluation, and demonstrates LLMs' scalability in few-shot learning. \textit{The Biases of Pre-Trained Language Models} \cite{biases} explores how prompt engineering affects sentiment analysis performance, identifying biases in label-word selection and fine-grained emotion taxonomies, with RoBERTa consistently outperforming other PLMs in classification accuracy.

\textit{RVISA: Reasoning and Verification for Implicit Sentiment Analysis} \cite{RVISA} significantly outperforms state-of-the-art (SOTA) baseline methods in implicit sentiment tasks, using accuracy and macro-F1 score as evaluation metrics. The paper also introduces new methods of fortifying training methods with more accurate material. Training the ED model with refined rationales generated by the DO GPT-3.5-turbo results in improved reasoning ability for the ED Flan-T5 model. The TH-RE approach helps justify the outputs generated, as the reasoning and outcome is derived in a step-by-step manner. The added level of verification makes the results more reelable, and allows the model to train on more accurate data. The author also suggests that the TH-RE method could also be applied to other SA applications, such as sarcasm detection.
adjustments. 

\subsection{Personality Recognition and Multimodal Emotion Detection}

\textit{'VyaktitvaNirdharan: Multimodal Assessment of Personality and Trait Emotional Intelligence'} \cite{VyaktitvaNirdharan} is a novel effort towards personality recognition and Trait EI measurement. The authors emphasize the direct relation between personality and EI, and hence develop a multitasking model that predicts both personality and EI trait. The authors assert that this is the first Hindi-based personality assessment system, making it a significant contribution to the field by expanding research into diverse cultural contexts. It also makes use of a novel dataset, which derives data from near real-time conversations with minimal intervention, and makes use of self-rated labels for reducing labeling bias. The study also demonstrated practical implementation by deploying it in two separate fields (recruiting and psychology).

Is ChatGPT a Good Personality Recognizer? Paper \cite{personality-recognizer} underlines the effectiveness of ChatGPT for zero-shot CoT prompting and surpasses significantly conventional RNN and fine-tuned RoBERTa methods for personality prediction. KnowleNet: Knowledge Fusion Network for Multimodal Sarcasm Detection approach \cite{knowleNet} fuses the common-sense knowledge for detecting sarcasm, which brought huge improvement of semantic consistency among inputs of text and image, outperforming current state-of-the-art with generalization on more datasets.

\subsection{Explainability and Ethical AI}

The paper \textit{Generative Ghosts: Anticipating Benefits and Risks of AI Afterlives}\cite{generativeghosts} has a timely and thought-provoking exploration of a novel concept. It contributes a useful design space framework that can guide future system development and policy-making. The authors offer a balanced and well-structured ethical analysis, thoughtfully weighing both the potential benefits and societal risks of posthumous AI agents. The usage of real-world examples and emerging commercial tools adds relevance and practical insight.

\textit{SenticNet 7: A Commonsense-Based Neurosymbolic AI Framework for Explainable Sentiment Analysis} \cite{SenticNet} provides an interpretable sentiment classification model by combining symbolic reasoning with deep learning. The system eliminates the need for labeled data while maintaining domain independence, making it a powerful alternative to traditional black-box models. In contrast, 	\textit{The Biases of Pre-Trained Language Models} \cite{biases} identifies ethical concerns in sentiment classification, advocating for bias mitigation in affective computing applications.

\subsection{Applications in Therapy and Safety}
\textit{'EmoTake: Exploring Drivers’ Emotion for Takeover Behavior Prediction'} is a novel study on implementing safety in semi-automated vehicles. The data used in this study has been derived from scratch by the authors, making it a valuable addition to the dataset library for this field. Furthermore, the input from the user is taken from a single camera, rather than multiple expensive and obstructive wearable biosensor devices. The developed framework showed promising results with high accuracy in multiple areas of prediction. 

\textit{The "Conversation" about Loss} \cite{grief-chatbot} explores chatbot technology for grief support, showcasing their role in emotional processing. Meanwhile, \textit{EMMA: An Emotion-Aware Wellbeing Chatbot} \cite{emma} automates affect detection, offering personalized responses for mental health interventions. 	\textit{An Affectively Aware Virtual Therapist} \cite{virtual-therapist} extends this work by integrating speech and facial emotion recognition, improving real-time emotional engagement in counseling.

Affective Computing in Safety Technologies
\textit{Predicting Driver Self-Reported Stress by Analyzing the Road Scene} \cite{driver-stress} presents a novel application of affective computing, predicting driver stress using visual road scene data. The study integrates spatial, object-based, and temporal features, showcasing the feasibility of stress-aware intelligent vehicle systems.

\subsection{Sarcasm Detection}
\textit{Sarcasm Detection in News Headlines using Supervised Learning} \cite{sarcasm-news} evaluates several linguistic models for the task of sarcasm detection. It is shown how RoBERTa outperforms more naturalistic models in capturing subtle contextual cues. The paper highlights the importance of bidirectional semantics in sarcasm detection tasks.

\section{Datasets for Emotion Management and Sentiment Analysis} 
The foundation of any successful affective computing system is the quality of the data on which it is trained. Emotion recognition, sentiment analysis, and personality assessment all rely heavily on annotated datasets that reflect the variation and complexity of human emotional expression. In the surveyed literature, researchers have utilized a wide variety of datasets sourced from different modalities and sources. Some of these studies leverage well-known benchmark datasets, while others present new sources to accomplish specific affective computing tasks. This section categorizes the datasets used in two broad categories - Text-based and Multimodal, and highlights their relevance, scope, and application in each study. Table \ref{dataset-summary-1} summarizes the datasets used in different research works studied in this review.

\begin{table*}[htbp]
\centering
\caption{Categorization of Datasets Used in Surveyed Papers}
\begin{tabular}{|p{4cm}|p{3cm}|p{3cm}|p{3cm}|p{3cm}|}
\hline
\textbf{Paper Title} & \textbf{Dataset Name} & \textbf{Category} & \textbf{Language} & \textbf{Usage} \\
\hline
\textit{RVISA: Reasoning and Verification for Implicit Sentiment Analysis}\cite{RVISA} & SemEval-2014 Restaurant Dataset\cite{semeval-restaurants}, SemEval-2014 Laptop Dataset\cite{semeval-laptops} & Text-Based & English & To generate rationales for each input sentence, for the multi-task fine-tuning stage. \\
\hline
\textit{Novel Speech-Based Emotion Climate Recognition in Peers’ Conversations Incorporating Affect Dynamics and Temporal Convolutional Neural Networks}\cite{peers} & K-EmoCon\cite{KEmoCon} & Multimodal – Physiological Signals & Korean \& English & Emotion climate recognition using physiological signals. \\
\cline{2-5}
 & The Interactive Emotional Dyadic Motion Capture (IEMOCAP)\cite{IEMOCAP} & Multimodal – Audio, Video & English & Emotion recognition in dyadic conversations. \\
\cline{2-5}
 & SEWA\cite{SEWA-paper}\cite{SEWA-site} & Multimodal – Audio, Video & Includes English, German, Hungarian, Greek, Serbian, and Chinese & Cross-cultural, continuous emotion recognition and sentiment analysis \\
\hline
\textit{GPTEval: A survey on assessments of ChatGPT and GPT-4}\cite{GPTEval} 
& GLUE, SuperGLUE\cite{superglue} & Text-Based & \multirow{7}{*}{English} & Language Proficiency \\
\cline{2-3} \cline{5-5}
& XSum, WikiBio\cite{xsum}\cite{wikibio} & Text-Based & & Summarization and Structured Generation \\
\cline{2-3} \cline{5-5}
& MMLU, GSM8K\cite{mmlu}\cite{gsm8k} & Text-Based & & Reasoning and Logic \\
\cline{2-3} \cline{5-5}
& PubMedQA, MedQA\cite{pubmedqa}\cite{medqa} & Text-Based & & Scientific and Medical QA \\
\cline{2-3} \cline{5-5}
& CodeXGLUE\cite{codexglue} & Text-Based & & Code Generation and Understanding \\
\cline{2-3} \cline{5-5}
& StereoSet, CrowS-Pairs\cite{stereoset}\cite{crows} & Text-Based & & Social Bias and Ethics \\
\cline{2-3} \cline{5-5}
& SST2, EmotionX, Reddit\cite{sst2}\cite{emotionx} & Text-Based & & Emotion and Sentiment Analysis \\
\hline
\textit{VyaktitvaNirdharan: Multimodal Assessment of Personality and Trait Emotional Intelligence\cite{VyaktitvaNirdharan}} & Vyaktitv dataset & Multimodal – Audio, Video & Hindi & Multimodal personality and trait EI assessment from Hindi dyadic conversations.\\
\hline
\textit{EmoTake: Exploring Drivers’ Emotion for Takeover Behavior Prediction}\cite{emotake} 
& EmoTake Dataset\cite{emotakedataset}
& Multimodal – Video, Other 
& Chinese / English 
& Driver emotion-based prediction of takeover readiness, reaction time, and quality using vision-based cues and vehicle data \\
\hline
\textit{Is ChatGPT a Good Sentiment Analyzer? A Preliminary Study}\cite{sentiment-analyzer}
& SST-2\cite{sst2} & Text-Based & \multirow{5}{*}{English} & Binary sentiment classification \\
\cline{2-3} \cline{5-5}
& SemEval 2014 Restaurant\cite{semeval-restaurants} & Text-Based & & Aspect-based sentiment analysis (ABSA) \\
\cline{2-3} \cline{5-5}
& SemEval 2014 Laptop\cite{semeval-laptops} & Text-Based & & Aspect-based sentiment analysis (ABSA) \\
\cline{2-3} \cline{5-5}
& Camera Dataset & Text-Based & & Comparative sentiment analysis (CSI, CEE tasks) \\
\cline{2-3} \cline{5-5}
& Emotion Cause Dataset \cite{emotion-cause-dataset}& Text-Based & & Emotion-cause and emotion-cause pair extraction \\
\hline
\textit{Is ChatGPT a Good Personality Recognizer? A Preliminary Study}\cite{personality-recognizer}
& Essays Dataset\cite{essays-dataset} & Text-Based & English & Personality trait prediction from structured, self-reflective essays \\
\cline{2-5}
& PAN Dataset\cite{pan-dataset} & Text-Based & English, Spanish, Italian and Dutch & Personality classification from noisy, real-world social media text (tweets) \\
\hline
\textit{KnowleNet: Knowledge Fusion Network for Multimodal Sarcasm Detection}\cite{knowleNet} 
& Multimodal Sarcasm Detection Dataset (Twitter)\cite{sarcasm-twitter} & Multimodal – Text, Image & English & Benchmarking sarcasm detection performance with textual, visual, and semantic cues \\
\cline{2-5}
& MultiBully (Twitter/Reddit)\cite{sarcasm-twitred} & Multimodal – Text, Image & English & Evaluating model generalizability in sarcasm detection across platforms \\
\hline
\textit{Is ChatGPT a General-Purpose Natural Language Processing Task Solver?}\cite{nlp-solver}
& GSM8K\cite{gsm8k}, CSQA\cite{csqa}, COPA\cite{copa} & Text-Based & English & Arithmetic, commonsense, symbolic, and logical reasoning \\
\cline{2-3}\cline{5-5}
& RTE\cite{rte}, CB\cite{cb} & Text-Based &  & Natural Language Inference (NLI) / entailment recognition \\
\cline{2-3}\cline{5-5}
& BoolQ\cite{boolq} & Text-Based &  & Binary QA for factive inference \\
\cline{2-3}\cline{5-5}
& MuTual\cite{mutual} & Text-Based &  & Dialogue reasoning in multi-turn conversations \\
\cline{2-3}\cline{5-5}
& SAMSum\cite{samsum} & Text-Based &  & Abstractive summarization of dialogue using ROUGE metrics \\
\hline
\end{tabular}
\label{dataset-summary-1}
\end{table*}

\begin{table*}[htbp]
\centering
\begin{tabular}{|p{4cm}|p{3cm}|p{3cm}|p{3cm}|p{3cm}|}
\hline
\cline{2-3}\cline{5-5}
\textit{Is ChatGPT a General-Purpose Natural Language Processing Task Solver?}\cite{nlp-solver} & CoNLL03\cite{conll03} & Text-Based & English & Named Entity Recognition (NER) with standard entity types \\
\cline{2-3}\cline{5-5}
& SST-2\cite{sst2} & Text-Based &  & Binary sentiment classification \\
\hline
\textit{A Wide Evaluation
of ChatGPT on Affective Computing Tasks}\cite{gpt-wide-eval}
& SemEval 2014, 2015 (Laptop, Restaurant)\cite{semeval-restaurants} \cite{semeval-laptops} & Text-Based &  & Aspect extraction, opinion identification, and polarity classification for ABSA \\
\cline{2-3}\cline{5-5}
& Twitter140\cite{twitter140} & Text-Based &  & Sentiment classification of tweets (positive/negative) \\
\cline{2-3}\cline{5-5}
& SemEval 2017 Task 5 \cite{semeval-2015} & Text-Based & English & Sentiment intensity ranking using [-1, 1] scores on microblogs and headlines \\
\cline{2-5}
& WASSA-2017 Emotion Intensity \cite{wassa-2017} & Text-Based &  & Emotion intensity regression (joy, fear, sadness, anger) using 0–1 scores \\
\cline{2-3}\cline{5-5}
& Reddit (Depression, Teenagers) \cite{teenage-reddit} & Text-Based &  & Suicide tendency detection via binary classification \\
\cline{2-3}\cline{5-5}
& Toxic Comment Classification Challenge \cite{toxic} & Text-Based & English & Multilabel toxicity detection (e.g., threats, insults, identity hate) \\
\cline{2-3}\cline{5-5}
& Reddit, Twitter (Well-being posts) \cite{well-being} & Text-Based & & Stress/well-being classification with binary labels \\
\cline{2-3}\cline{5-5}
& Kaggle TEDTalks Tweets \cite{tedtalks-kaggle} & Text-Based & & Engagement prediction using log-transformed retweet counts \\
\cline{2-3}\cline{5-5}
& First Impressions Dataset \cite{first-impressions} & Text-Based & & Big Five personality prediction (regression task in [0, 1]) \\
\cline{2-3}\cline{5-5}
& Sarcasm Headlines Dataset (The Onion, HuffPost) \cite{onion-huff} & Text-Based & & Binary classification of sarcastic vs. non-sarcastic news headlines \\
\cline{2-3}\cline{5-5}
& Rotten Tomatoes Reviews \cite{rotten-tomatoes} & Text-Based & & Binary subjectivity classification (subjective vs. objective) \\
\hline
\textit{Sentiment Analysis in the Era of Large Language Models: A Reality Check}\cite{sentiment-llm}
& IMDb\cite{imdb}, Yelp-2, Yelp-5\cite{yelp-2-5}, Movie Reviews\cite{MR}, SST2\cite{sst2}, Twitter\cite{twitter-sentiment-llm}, SST5\cite{sst5}, SemEval 2014, 2015 (Laptop, Restaurant)\cite{semeval-restaurants} \cite{semeval-laptops} & Text-Based & English & Sentiment classification and aspect-based sentiment structure extraction.\\
\cline{2-3}\cline{5-5}
& HateEval\cite{hateeval}, Irony18\cite{irony18}, OffensEval\cite{offenseval}, Stance16\cite{stance16}, CS19\cite{cs19}, Emotion20\cite{emotion20} & Text-Based & & Multifaceted subjective text analysis (hate, irony, stance, emotion, etc.)\\
\hline
\textit{The "Conversation" about Loss: Understanding How Chatbot Technology was Used in Supporting People in Grief}\cite{grief-chatbot} 
& Chatbot Grief Support Interviews 
& Text-Based
& English 
& Semi-structured interviews and chat logs with mourners, thematically analyzed for emotional and interactional patterns. \\
\hline
\textit{Will Affective Computing Emerge from Foundation Models and General AI? A First Evaluation on ChatGPT}\cite{first-gpt}
& First Impressions Dataset \cite{first-impressions} & Text-Based & & Predict Big Five personality traits from video transcriptions using regression and binarization. \\
\cline{2-3}\cline{5-5}
& Sentiment140\cite{twitter140} & Text-Based & English & Binary sentiment classification of tweets (neutral class filtered); tests model robustness on noisy data. \\
\cline{2-3}\cline{5-5}
& Reddit (SuicideWatch, Depression, Teenagers) \cite{dep-detection} & Text-Based & & Classify suicidal and depressive tendencies in Reddit posts to evaluate model effectiveness in detecting extreme emotional states. \\
\hline
\textit{Sarcasm Detection in News Headlines using Supervised Learning}\cite{sarcasm-news} 
& Sarcasm Headlines Dataset (The Onion, HuffPost) \cite{onion-huff} & Text-Based & English & Binary sarcasm classification using cleaned news headlines; two dataset versions evaluated after train/val/test split \\
\hline
\textit{The Biases of Pre-Trained Language Models: An Empirical Study on Prompt-Based Sentiment Analysis and Emotion Detection}\cite{biases}
& Amazon Product Reviews (Books, DVDs, Electronics, Kitchen) \cite{amazon-reviews} & Text-Based & & Binary sentiment classification using prompts to evaluate model bias in polarity prediction \\
\cline{2-3}\cline{5-5}
& WASSA-2017 Emotion Intensity Dataset\cite{wassa-2017} & Text-Based & English & Emotion classification (anger, fear, joy, sadness) using cleaned tweets; tested prompt-based model performance \\
\hline
\end{tabular}
\label{dataset-summary-2}
\end{table*}

\begin{table*}[htbp]
\centering
\begin{tabular}{|p{4cm}|p{3cm}|p{3cm}|p{3cm}|p{3cm}|}
\hline
\textit{SenticNet 7: A Commonsense-based Neurosymbolic AI Framework for Explainable Sentiment Analysis}\cite{SenticNet}
& Customer Reviews\cite{customer-reviews}, Movie Reviews\cite{MR}, Amazon Reviews \cite{amazon-reviews}, IMDb \cite{imdb}& Text-Based &  & Binary sentiment classification of product reviews \\
\cline{2-3}\cline{5-5}
& Sanders \cite{sanders}, SemEval 2015, 2016)\cite{semeval-restaurants} \cite{semeval-laptops} & Text-Based & English & Sentiment classification of brand-related tweets \\
\cline{2-3}\cline{5-5}
& SST \cite{sst2} & Text-Based & & Sentence-level sentiment classification \\
\cline{2-3}\cline{5-5}
& STS \cite{sts} & Text-Based & & Polarity and subjectivity classification of sentences \\
\cline{2-3}\cline{5-5}
& SemEval 2013 & Text-Based & & Sentiment analysis of tweets.\\
\hline
\textit{Predicting Driver Self-Reported Stress by Analyzing the Road Scene}\cite{driver-stress} 
& AffectiveROAD\cite{affectiveroad-dataset} 
& Multimodal – Video 
& English (visual scenes) 
& Predict drivers’ stress from road scene videos using human-annotated stress scores (0–1), discretized into low, medium, and high classes \\
\hline
\textit{EMMA: An Emotion-Aware Wellbeing Chatbot}\cite{emma}
& Smartphone Sensor Data & Multimodal – Geolocation, Contextual Features & English & Used to infer valence and arousal via personalized ML models based on passive geolocation and time-based features \\
\cline{2-5}
& Experience Sampling Data & Text-Based & English & Ground-truth emotional labels collected via Russell’s circumplex model to train and validate mood predictions \\
\cline{2-5}
& Psychometric Survey Data (Big Five, PANAS, DASS) & Text-Based (Survey) & English & Captures personality and mood baselines; used for personalization and context in affective modeling \\
\hline
\textit{An Affectively Aware Virtual Therapist for Depression Counseling}\cite{virtual-therapist}
& Emotional Prosody Speech and Transcripts (LDC) \cite{speech-transcript}& Multimodal – Audio, Text & English & Used OpenSmile features to classify speech into happy, neutral, and sad categories for affect detection training \\
\cline{2-5}
& Affdex SDK Facial Emotion Data\cite{amfed} & Multimodal – Video & English & Facial emotion recognition system used in parallel with speech input to maintain affect awareness \\
\cline{2-5}
& Pilot Study Dataset (10 Participants) & Multimodal – Audio, Text, Survey & English & User interactions and feedback used to evaluate satisfaction, trust, and engagement with the virtual agent \\
\hline
\end{tabular}
\label{dataset-summary-2}
\end{table*}

\subsection{Text-Based Datasets}

\textit{RVISA: Reasoning and Verification for Implicit Sentiment Analysis} uses the \textbf{Restaurant Dataset (SemEval-2014)} and \textbf{Laptop Dataset (SemEval-2014)}. The Restaurant dataset contains reviews of restaurants in areas such as food, service and ambiance. The sentiment polarity of these aspects is labeled, which models use to learn for predicting sentiments related to different aspects of the dining experience. Similarly, the Laptop dataset consists of reviews on laptops. In this case, each review includes aspect terms relevant to laptops (e.g., battery life, performance, design), and the sentiment polarity for these aspects is also labeled.

The paper \textit{GPTEval: A Survey on Assessments of ChatGPT and GPT-4} reviews various datasets utilized in evaluating ChatGPT and GPT-4 across multiple tasks. Regarding language proficiency, it employs \textbf{GLUE, SuperGLUE, XSum, and WikiBio} to evaluate general NLP capabilities. All the above is measured using the following benchmarks: MMLU (Massive Multitask Language Understanding) and GSM8K for reasoning skills with arithmetic and logical tasks; scientific knowledge on PubMedQA; MedQA for medical reasoning; CodeXGLUE for coding-related tasks; and StereoSet and CrowS-Pairs for social understanding and ethical consideration, focusing on bias detection. The capability of the models in affective computing is portrayed through the analysis of sentiment and emotion with the \textbf{SST2, EmotionX and Reddit} dataset.

\textit{Is ChatGPT a Good Sentiment Analyzer? A Preliminary Study} evaluates the performance of ChatGPT on Sentiment Analysis with several benchmark datasets\cite{sentiment-analyzer}. SST-2 will be used for binary sentiment classification; since the test set is not available, the validation set is utilized and this dataset has been the standard benchmark for general sentiment analysis. The SemEval 2014 ABSA Challenge Datasets (14-Restaurant, 14-Laptop) focus on the ABSA task in which a model can classify sentiment polarity depending on particular aspects in customer reviews, thus assessing its capability in fine-grained sentiment analysis. The Camera Dataset is used for the tasks of CSI and CEE; it helps extract and analyze sentiment from comparative statements, including subject-object relationships and evaluative aspects. Finally, \textbf{Emotion Cause Dataset} is utilized for the tasks of \textbf{Emotion Cause Extraction (ECE) and Emotion-Cause Pair Extraction (ECPE)}-enabling the model to find out emotion triggers and pair them with their causes to deeply comprehend semantic and emotional causality in sentiment-driven text.

\textit{Is ChatGPT a Good Personality Recognizer? A Preliminary Study} used two datasets for the evaluation of personality recognition\cite{personality-recognizer}. The \textbf{Essays Dataset} consists of 2,467 essays from psychology students, in which personality traits are measured using self-tagging questionnaires on traits such as Openness, Conscientiousness, Extraversion, Agreeableness, and Neuroticism. The data is structured and self-reflective, therefore rich in personal context. The \textbf{PAN Dataset} comprises tweets from 294 users; their personality traits have been binarized into high and low scores based on the mean scores. This dataset introduces real-world, noisy, and user-generated content, assessing the model’s adaptability across diverse linguistic styles. Both datasets were split into training (80\%), validation (10\%), and testing (10\%) to ensure robust evaluation.

\textit{Is ChatGPT a General-Purpose Natural Language Processing Task Solver?} discusses zero-shot performance on 20 datasets categorized into seven categories of NLP tasks \cite{nlp-solver}. Reasoning includes arithmetic reasoning, commonsense, symbolic reasoning, and logical reasoning, and its assessment by the datasets \textbf{GSM8K}, \textbf{CSQA}, and \textbf{COPA} with multi-step inferencing for the prompt given. \textbf{NLI: RTE} and \textbf{CB} datasets were employed for entailment recognition. \textbf{QA: BoolQ} dataset was used for binary entailment on factive inference. \textbf{MuTual} was utilized to test the model's prowess in understanding conversations in multiple turns. \textbf{SAMSum} for abstractive dialogue summarization is quantified with ROUGE scores. \textbf{NER: CoNLL03} for categorical entities like Person, Organization, and Location. Finally, the \textbf{Sentiment Analysis} was evaluated using \textbf{SST2} for binary sentiment classification. All these datasets test ChatGPT on generalization and reasoning in a wide variety of NLP domains.

\textit{A Wide Evaluation of ChatGPT on Affective Computing Tasks} tests the capabilities of ChatGPT in affective computing based on several datasets\cite{gpt-wide-eval}. For ABSA, the datasets used were \textbf{SemEval 2014} and \textbf{SemEval 2015} (laptop and restaurant reviews) for evaluating aspect extraction and polarity classification. General sentiment classification is done using Twitter140, while the testing of the sentiment intensity ranking is done with SemEval-2017 Task 5 (microblogs and news headlines). The intensity of the emotion is measured by WASSA-2017 on a range of joy, sadness, fear, and anger. In the analysis of mental health, Reddit posts originating from depression-related subreddits support suicide tendency detection, while the Toxic Comment Classification Challenge-Wikipedia comments-assist in toxicity detection by identifying offensive traits. Well-being assessment on \textbf{Reddit} and \textbf{Twitter} data: to identify stress-positive versus stress-negative content. \textbf{Kaggle TEDTalks Tweets} is used in measuring engagement: this dataset was collected to study the virality of tweets. Personality assessment: the \textbf{First Impressions} dataset comprises of video transcripts annotated with Big Five personality traits. Sarcasm detection: news headlines from \textbf{The Onion} and \textbf{HuffPost}; subjectivity detection: \textbf{Rotten Tomatoes} movie reviews for subjective versus objective content classification. The original data splits were maintained when possible; otherwise, the data was split into training, validation, and test sets for consistency.

Sentiment Analysis in the Era of Large Language Models: A Reality Check by \cite{sentiment-llm} measured the performance of sentiment analysis on 13 tasks using 26 datasets, divided into three broad areas.
For SC, the datasets are IMDb, Yelp-2, Yelp-5, MR, SST2, Twitter, SST5, Lap14, and Rest14. Perform binary/multi-class/aspect-specific sentiment classification. Sentiment labels like positive, negative, and neutral are annotated on different levels of documents/sentences/aspects. Accuracy serves as the evaluation metric for these datasets.
For the \textbf{ABSA} task, Unified Aspect-Based Sentiment Analysis (UABSA), Aspect Sentiment Triplet Extraction (ASTE), and Aspect Sentiment Quadruple Prediction (ASQP) were conducted on the \textbf{Laptop14, Rest14, Rest15, and Rest16} datasets. In these datasets, structured sentiment elements, including aspect, opinion, and polarity, were extracted for which evaluation is performed using the Micro-F1 score.
For MAST, implicit sentiment, hate speech detection, irony detection, offensive language identification, stance detection, comparative opinion mining, and emotion recognition were performed using datasets such as HateEval, Irony18, OffensEval, Stance16, CS19, and Emotion20. These datasets were evaluated based on the Macro-F1 score to measure nuanced sentiment understanding.

\textit{The "Conversation" about Loss: Observing How Chatbot Technology Was Deployed to Assist People in Mourning} grief-chatbot recruited participants through semi-structured interviews with 10 grievers who had used chatbots while mourning. The study also reused chat logs by analyzing those records.

Will Affective Computing Emerge from Foundation Models and General AI? A First Evaluation on ChatGPT \cite{first-gpt} employed several datasets for personality prediction, sentiment classification, and mental health assessment. For \textbf{Personality Recognition}, the \textbf{First Impressions (FI)} dataset was utilized to predict Big Five personality traits: Openness, Conscientiousness, Extraversion, Agreeableness, and Neuroticism. The video transcriptions obtained from FI, which were the regression values of personality labels in the range [0,1] and binarized at threshold 0.5 to represent positive or negative traits. For \textbf{Sentiment Analysis}, the \textbf{Sentiment140} dataset was adopted, consisting of labeled tweets. This dataset has been prepared for sentiment classification by filtering out the neutral tweets in order to create a binary classification. Finally, for the task of \textbf{Mental Health Detection,} the \textbf{Suicide and Depression Dataset} was composed of text posts from subreddits such as \textit{SuicideWatch} and \textit{Depression} (positive cases), while the ones on \textit{Teenagers} were labeled as negative. Posts with greater than 512 characters were removed to make each sample consistent.

\textit{Sarcasm Detection in News Headlines using Supervised Learning} \cite{sarcasm-news} uses the \textbf{News Headlines Dataset} in two versions for the purpose of classifying sarcasm.
Version 1 includes 26,709 headlines (\textit{11,724 sarcastic} and \textit{14,985 non-sarcastic}), while Version 2 includes 28,619 headlines (\textit{13,634 sarcastic} and \textit{14,985 non-sarcastic}). The sarcastic headlines were extracted from \textit{The Onion}, whereas the non-sarcastic headlines were collected from \textit{HuffPost}. Further, the dataset was split into training (\textit{80\%}), validation (\textit{10\%}), and testing (\textit{10\%}), and pre-processing steps such as Unicode correction, contraction expansion, case normalization, and punctuation removal were followed.

The Biases of Pre-Trained Language Models: An Empirical Study on Prompt-Based Sentiment Analysis and Emotion Detection by \cite{biases} uses two datasets for sentiment analysis and emotion detection. The dataset, Amazon Product Reviews Dataset, consists of customer reviews on books, DVDs, electronics, and kitchen appliances which are categorized as positive or negative sentiments. This data was pre-processed to fit within the token length constraints of pre-trained language models. Emotion detection was performed on the \textbf{WASSA-2017 Shared Task Dataset} with tweets annotated for \textit{anger, fear, sadness, and joy}. Text preprocessing included removing emojis, hashtags, and user IDs. Models were compared based on their performance in terms of the accuracy of fine-grained emotion classification.

\textit{SenticNet 7: A Commonsense-Based Neurosymbolic AI Framework for Explainable Sentiment Analysis} \cite{SenticNet} reports the results of its sentiment analysis on ten general datasets. One hundred. These include CR (Customer Reviews) MR (Movie Reviews) Amazon: reviews on products IMDb: movie reviews Sanders for brand-related tweet sentiment analysis SST Stanford Sentiment Treebank STS - Sentiment Treebank with Gold Annotations and finally, the SemEval SE13 SE15 SE16 challenges. These datasets assess \textit{binary sentiment classification} (positive/negative), and the work places SenticNet 7 in comparison to 20 sentiment lexica created between 1966 and 2020. All dataset labels were reduced to binary classes for consistency in evaluation.

\subsection{Multimodal Datasets}

\textit{Novel Speech-Based Emotion Climate Recognition in Peers’ Conversations Incorporating Affect Dynamics and Temporal Convolutional Neural Networks} has used a diverse and some of the most unique multimodal dataset, that is valuable to the field of Affective Computing \cite{peers}. \textbf{K-EmoCon} is a multimodal dataset that recorded the emotions of participants using biosensors who were indulging in a debate on rehabilitating Yemeni refugees in Jeju Island in South Korea. Their physiological signals (heart rate, EEG, body temperature, etc.).15 minutes after the conversation ended the participants were made to look back on the conversation and self-rate their emotions, along with ratings for their conversational partners. Next, \textbf{IEMOCAP}, a widely used actor-based dataset containing controlled conversations from ten unique speakers was also used. It had data in the form of audio and video, with each dialogue divided into utterances and annotated with six categorical emotions (happy, sad, neutral, angry, excited, frustrated) and three-dimensional emotion labels (arousal, valence, domination). Lastly, \textbf{SEWA} is a cross-cultural dataset that captures emotions during interactions across six languages, namely - British English, German, Hungarian, Greek, Serbian, and Chinese. The participants' audio-visual behaviors were recorded and the footage has been annotated for valence and arousal by external annotators.

The paper \textit{'VyaktitvaNirdharan: Multimodal Assessment of Personality and Trait Emotional Intelligence'} uses a self-generated dataset. The researchers recorded multiples sets of conversations between two individuals from a pool of selected participants using microphones and cameras. After this process of minimal disruption for collection of data, participants self-rated themselves by answering a personality assessment. It was measured using the Goldberg’s '100 Unipolar Markers' and were mapped to one of the following: Openness, Conscientiousness, Extroversion, Agreeableness, and Neuroticism. Trait EI was measured using the TEIQue-SF questionnaire, which evaluates the participant under the following areas: Well-being, Self-control, Sociability, Emotionality. As the personality and EI assessment did not involve an external rater, it eliminated the labeling bias. The dataset was processed to extract visual, acoustic, and linguistic features. For the visual feature attraction, the OpenFace 2.0 toolkit was used for the visual feature processing. Key frame extraction was performed using the efficient LUV color space variation. For audio feature processing, the LibROSA package was used. More particularly, the fundamental frequency (F0) assisted in tone analysis. MFCCs was employed for speech quality and phonetic structure. Energy levels of voice was also recorded to assess vocal intensity.

The dataset for EmoTake \cite{emotake} was collected using a controlled driving simulator experiment. A high-fidelity Level 3 automated vehicle simulation was used, where drivers were asked to watch different selected videos as NDRTs before being issued a TOR. A multimodal data collection approach was employed, using computer vision-based emotion and movement analysis tools. OpenFace 2.0, a state-of-the-art facial behavior analysis toolkit, was used to track facial expressions, head posture, and gaze movement, to help the model with precise emotion recognition. For tracking the upper-body posture, OpenPose 2.0, a deep-learning-based pose estimation framework, was utilized.

\textit{KnowleNet: Knowledge Fusion Network for Multimodal Sarcasm Detection} utilizes two multimodal sarcasm detection datasets sourced from social media\cite{knowleNet}. \textbf{Dataset 1} consists of 19,816 samples from Twitter, incorporating text, images, image attributes (keywords), and captions. It is split into training (8,642 sarcastic, 11,174 non-sarcastic), development (959 sarcastic, 1,451 non-sarcastic), and test (959 sarcastic, 1,450 non-sarcastic) sets. \textbf{Dataset 2} comprises 5,854 samples from Twitter and Reddit, containing text and images with sarcasm labels. The datasets were used to benchmark KnowleNet against state-of-the-art unimodal and multimodal models, with Dataset 2 evaluating its generalizability. It used image captioning for preprocessing with MobileNetV3 and encoding using ResNet. For textual features, BERT embeddings were used.

\textit{Predicting Driver Self-Reported Stress by Analyzing the Road Scene} \cite{driver-stress} uses the \textbf{AffectiveROAD} dataset to predict drivers' self-reported stress based on videos of the road scene. The dataset is made up of 676.3 minutes of synchronized video recordings annotated with regards to stress levels on a 0-1 scale, representing subjective perceptions of stress. This work categorizes the levels into three classes, namely: \textit{low stress} (\(0-0.4\)), \textit{medium stress} (\(0.4-0.75\)), and \textit{high stress} (\(0.75-1\)). Accordingly, the normalized and discretized classes from given stress labels for classification were used.

\textit{EMMA: An Emotion-Aware Wellbeing Chatbot} \cite{emma} utilizes three data sources for emotion inference: \textbf{sensor data}, \textbf{experience sampling}, and \textbf{psychometric surveys}. \textbf{Sensor data} includes passive smartphone-collected \textit{geolocation}, \textit{derived features} (e.g., distance from home/work), and \textit{contextual information} (e.g., time of day). This data was transformed into high-level features for emotion prediction. \textbf{Experience sampling data} served as ground-truth labels, where participants self-reported their emotional states based on \textit{Russell’s circumplex model of emotion}. \textbf{Psychometric surveys} included the \textit{Big Five Personality Test} for trait modeling, \textit{PANAS} (Positive and Negative Affect Schedule) for mood quantification, and \textit{DASS} (Depression, Anxiety, and Stress Scales) for initial mental health assessment.

\textit{An Affectively Aware Virtual Therapist for Depression Counseling} \cite{virtual-therapist} integrates multimodal datasets to detect user emotions. The affect detection system was trained on Emotional Prosody Speech and Transcripts from the Linguistic Data Consortium for classifying speech into happy, neutral, and sad valence categories, each with approximately 160 samples. Features were extracted using OpenSmile and then classified using LibSVM. Facial Emotion Recognition via Affdex SDK: This module runs in parallel to the speech-based detection to continuously recognize user expressions. A pilot study dataset of 10 participants diagnosed with mild to moderate depression, PHQ-9 scores between 5 and 14, provided real-world user-interaction data while collecting audio recordings, emotional feedback, and engagement metrics, assessing trust and user satisfaction.

\section{Evaluation Methodologies}
The testing and evaluation of discussed affective computing applications employ different approaches. Each of these methods aim to measure their accuracy, effectiveness and give an insight on where it stands compared to related work in that area of research.

This section describes some of the standard evaluation metrics and procedures used for Affective Computing applications. Next, some of the effective and innovative the evaluation methodologies used in the studied papers are explained. Table \ref{standard-evaluation-methods} lists and briefly explains standard evaluation metrics and procedures in affective computing.

\begin{table*}[htbp]
\centering
\caption{Standard Evaluation Metrics and Procedures in Affective Computing}
\begin{tabular}{|p{3.5cm}|p{5.5cm}|p{6.5cm}|}
\hline
\textbf{Category} & \textbf{Metric / Method} & \textbf{Purpose / Notes} \\
\hline
\textbf{Classification} & Accuracy & Measures overall performance of ML models \\
\cline{2-3}
& Precision, Recall, F1-score (macro/micro/weighted)& Standard for multi-class emotion tasks \\
\cline{2-3}
& Confusion Matrix & Visualizes class-wise prediction performance \\
\cline{2-3}
& Unweighted Average Recall (UAR) & Useful when class distribution is imbalanced \\
\hline
\textbf{Regression} & Mean Absolute Error (MAE) / Root Mean Squared Error (RMSE) & Measures error in predicting continuous emotion variables (e.g., valence, arousal) \\
\cline{2-3}
& Pearson / Spearman Correlation & Measures alignment between predicted and ground truth emotion scores \\
\hline
\textbf{Multimodal \& Behavioral} & Area Under Curve (AUC) & Evaluates performance across thresholds. Popular for physiological emotion models. \\
\cline{2-3}
& Temporal Stability / Sequence Accuracy & For emotion recognition from video/audio over time \\
\cline{2-3}
& Engagement Metrics (Latency, Frequency) & Relevant for chatbots and interactive affective systems \\
\hline
\textbf{Validation Methods} & k-Fold Cross-Validation & Measures generalizability across data splits \\
\cline{2-3}
& Leave-One-Subject-Out (LOSO) & Tests robustness across different individuals \\
\cline{2-3}
& Ablation Studies & Identifies contributions of individual components \\
\cline{2-3}
& Baseline Comparisons & Benchmarks against classical and state-of-the-art models \\
\hline
\textbf{Human Evaluation} & Likability, Trust, Satisfaction (Likert Scales) & Subjective assessment in interactive or therapeutic settings \\
\cline{2-3}
& Interviews / Open-ended Feedback & Used for qualitative insights in affective support systems. A Likert scale is a psychometric tool commonly used in surveys, questionnaires, and human-centered research to measure perceived experiences. \\
\hline
\textbf{Fairness \& Explainability} & Demographic Bias Testing & Evaluates fairness across gender, age, etc. \\
\cline{2-3}
& Explainability Tools (e.g., GradCAM) & Visual tools to interpret model focus areas in emotion prediction \\
\hline
\end{tabular}
\label{standard-evaluation-methods}
\end{table*}

The evaluation techniques used for RVISA framework\cite{RVISA}, introduces a novel \textbf{answer-based verification step} which improves implicit sentiment classification and demonstrates the importance of multi-step prompting and rationale in performance and explainability mechanisms. The performance of RVISA was measured using \textbf{accuracy and macro-F1 score}, comparing it against baselines BERT and the THOR framework.

The AffECt framework\cite{peers} by G. Alhussein et al. was evaluated using datasets generated on the basis of \textbf{spontaneous conversations}. This increases the credibility of the system as the results show that the system is more likely to perform better in a real-world application. The model’s performance was assessed through \textbf{leave-one-sample-out cross-validation} demonstrating its generalizability and compared against several deep learning baselines and state-of-the-art methods.

In real-world applications, EMMA\cite{emma} and EmoTake\cite{emotake} are evaluated through \textbf{user studies}, showing their effectiveness with real participants. EMMA, an emotion-aware wellbeing chatbot, was tested in a \textbf{randomized controlled trial}, where both engagement metrics (e.g., latency, frequency) and perceived likability were tracked alongside ML performance. Similarly, EmoTake’s use of a \textbf{driving simulator study}, labeled by both users and experts, highlights the value of contextual and behavioral metrics in emotion-based interface evaluations.

SenticNet 7\cite{SenticNet} exemplifies scalable and explainable evaluation by \textbf{benchmarking against 20 sentiment lexica} without supervised fine-tuning. Ablation studies conducted on the neurosymbolic modules highlights the importance of explainability in affective computing.

\section{Challenges and Weaknesses}
Despite significant advancements in affective computing and sentiment analysis, the surveyed studies reveal several limitations and challenges. Key challenges include the dependency on high-quality labeled data, performance inconsistencies across domains, inherent biases in pre-trained language models, and limitations in handling nuanced or multimodal affective states. Additionally, the studies highlight concerns regarding ethical considerations, fairness, and trustworthiness in emotion-aware AI applications, particularly in sensitive domains such as mental health and well-being.

\subsection{Evaluation and Generalization Constraints}

In the RVISA model \cite{RVISA}, the authors mention that the answer-based verification mechanism which checks the output of the ED model could be further optimized or use a different method. The author also highlights a limitation in the field of SA, where many areas such as suicide tendency, hate speech and sarcasm detection remain unexplored. They state that adapting more generic evaluation and training methods, RVISA can be expanded to broader applications. Another limitation identified through this survey is that the developed model was tested on only two datasets, thereby restricting its effetiveness on a wider variety of ISA tasks and more real-world applications.

The AffECt framework \cite{peers} has shown promising results to assess emotion climate (EC) but the authors have pointed out that the model shows lower specificity in \textbf{valence classification}, as it depends on subtle contextual and cultural cues. Though regularization techniques and feature optimization were applied, there is room for further improvement. Cross-dataset and cross-language variations due to linguistic and cultural differences also posed challenges. To address this, transfer learning and domain adaptation strategies were explored and additional refinement is required to ensure robustness across diverse conversational datasets. \textbf{Lack of diverse datasets} also poses challenges, as it forces the model to train on a biased database. The authors conclude by highlighting areas that can be explored in this direction. Some suggestions include expanding the modality of datasets such as integration of biosignals and video-based emotion cues. A more robust and accurate model could be applied in areas such as group therapy.

Most of the works point out the deficiencies in current methods, question overfitting to specific datasets, and real-world generalization. There is also a lack of \textbf{standard benchmarking scale or mechanism} in the field. The paper \textit{GPTEval: A Survey on Assessments of ChatGPT and GPT-4} \cite{GPTEval} presents challenges in \textbf{domain-specific reasoning} and \textbf{ambiguity resolution} wherein it requires more sophisticated benchmarking methods and enlarged datasets that are more inclusive and fair.

In \textit{VyaktitvaNirdharan: Multimodal Assessment of Personality and Trait Emotional Intelligence} \cite{VyaktitvaNirdharan}, the personality labeling were discretized into binary categories based on mean or median values, which may not fully capture the \textbf{complexity} of individual personality traits. Future research could explore \textbf{regression-based approaches} for more precise assessments. Additionally, the dataset collection relies on usage of video and audio recording which raises \textbf{data security} and \textbf{privacy} concerns. The authors highlight the importance of data anonymization and compliance with ethical guidelines.

\textit{Is ChatGPT a Good Sentiment Analyzer? A Preliminary Study} \cite{sentiment-analyzer} notes that the study's evaluation lacks rigor due to uncertainty about whether ChatGPT’s pre-training data included test samples. Additionally, the study did not extensively explore prompt engineering and had limited model comparisons due to \textbf{restricted access to APIs}.
\textit{Is ChatGPT a Good Personality Recognizer? A Preliminary Study} \cite{personality-recognizer} suggests that personality recognition could be enhanced by training domain-specific LLMs on psychological datasets, as ChatGPT’s current capabilities remain limited.\textit{A Wide Evaluation of ChatGPT on Affective Computing Tasks} \cite{gpt-wide-eval} highlights the need for broader evaluation, suggesting additional research on context-aware sentiment analysis, multimodal sentiment tasks, and multilingual datasets.
\textit{Sentiment Analysis in the Era of Large Language Models: A Reality Check} \cite{sentiment-llm} highlights that \textbf{benchmark datasets} might not be adequate to capture the complexity of \textbf{real-world sentiment}. The work proposes further research on fine-tuning and optimization of prompts for consistency.
The paper \textit{Sarcasm Detection in News Headlines using Supervised Learning} \cite{sarcasm-news} limits its study to news headlines only, leaving research gaps to understand sarcasm in social media and conversational contexts. Further research is suggested by expanding the study to semi-supervised or unsupervised learning techniques.
\textit{Will Affective Computing Emerge from Foundation Models and General AI? A First Evaluation on ChatGPT} \cite{first-gpt} has a limited dataset size and manual data processing, which restricts its scalability.

\subsection{Computational and Resource Constraints}

Several works have pointed to the computational limitation preventing large-scale evaluations, various model comparisons, and extensive task explorations.
Is ChatGPT a General-Purpose Natural Language Processing Task Solver? The work of \\cite{nlp-solver} observes that large-scale dataset evaluations are \textbf{limited by computational costs} and prevent testing across more task categories. There is scope to continue exploring the zero-shot vs. few-shot learning capabilities of ChatGPT.
\textit{KnowleNet: Knowledge Fusion Network for Multimodal Sarcasm Detection} \cite{knowleNet} requires \textbf{external commonsense knowledge sources} (e.g., ConceptNet), increasing computational overhead. Additionally, integrating triplet loss and multimodal learning modules results in higher training and inference times.
\textit{The Biases of Pre-Trained Language Models: An Empirical Study on Prompt-Based Sentiment Analysis and Emotion Detection} \cite{biases} raises concerns about selecting optimal prompts for classification tasks. While ensemble learning could mitigate this issue, the study does not explore this approach in depth.

\subsection{Multimodal Affective Computing Limitations}

While there is promising advancement in multimodal approaches for emotion recognition, challenges still remain regarding dataset diversity, model generalization, and complexity.

\textit{A Wide Evaluation of ChatGPT on Affective Computing Tasks }\cite{gpt-wide-eval} recommends further extension of multimodal sentiment tasks with inclusion of textual, audio, and video data.

\textit{An Affectively Aware Virtual Therapist for Depression Counseling} \cite{virtual-therapist} incorporates speech and facial emotion recognition but is limited by a \textbf{small sample size}. The study suggests expanding dataset size and interaction durations to improve real-world applicability.
\textit{EMMA: An Emotion-Aware Wellbeing Chatbot} \cite{emma} relies on older machine learning models and was evaluated on a \textbf{small, non-diverse participant group}. The study suggests adopting recent deep learning advancements and testing on clinically diverse populations.

\subsection{Dataset and Demographic Limitations}

The generalizability of affective computing models is often constrained by the diversity and scale of datasets. Several studies highlight limitations in dataset representation and demographic coverage.

In \textit{VyaktitvaNirdharan: Multimodal Assessment of Personality and Trait Emotional Intelligence} \cite{VyaktitvaNirdharan}, though the paper has made a good effort to work with a unique demographic, it has limited the representation as the dataset by primarily focusing on Hindi-speaking individuals, which may affect its \textbf{ability to generalize} to users from diverse linguistic and cultural backgrounds. The dataset was derived from a \textbf{small sample size}. Future studies need to expand on this work with a larger and more diverse dataset. The authors claim self-reported personality assessments, to eliminate bias, but could still be affected by individual biases such as social desirability and subjective interpretation. In this case, including a third-party labeling process as additional data could increase reliability and fortify the results.

Although the study \textit{EmoTake: Exploring Drivers’ Emotion for Takeover Behavior Prediction} was innovative in recording fresh data as part of the study, it constituted of a  
\textbf{limited sample size (26 participants)} and diversity. The testing environment was controlled ensuring factors such as appropriate lighting conditions, to favor the camera capturing the driver's emotional state and posture data, thus lacking insight to show its usefulness with real-world drivers and road conditions. Further, the participants were exposed to a limited type of NDRT. The authors also highlighted a \textbf{privacy concern}, as users have to be observed through a camera to capture data. They also mention that a single camera may not be sufficient in this highly critical application.

\textit{Predicting Driver Self-Reported Stress by Analyzing the Road Scene} \cite{driver-stress} is limited by the narrow demographic representation of its driver population. Expanding the dataset to include diverse geographic regions and driving conditions would improve model generalization.
\textit{EMMA: An Emotion-Aware Wellbeing Chatbot} \cite{emma} was tested on a small and non-clinical population. Future work should evaluate its effectiveness in clinically diagnosed individuals.
\textit{An Affectively Aware Virtual Therapist for Depression Counseling} \cite{virtual-therapist} had a limited participant pool, restricting its findings to a small sample size.

The \textbf{lack of multilingual users and data} during training and testing of systems were observed across several studies. This strongly shows the need for \textbf{multilingual} and \textbf{multicultural datasets}. A system designed with a diverse user base in mind would show better accuracy when deployed to users coming from different backgrounds and regions.

\section{Ethical and Societal Considerations}
Large Language Models (LLMs), including ChatGPT, inherit biases from the datasets they are trained on. The Biases of Pre-Trained Language Models \cite{biases} highlights that PLMs exhibit disparities in sentiment classification and emotion detection, influenced by prompt selection, dataset composition, and label-word choices. Similarly, GPTEval \cite{GPTEval} notes ethical concerns in ChatGPT’s responses, particularly regarding fairness across different demographic groups and the potential reinforcement of stereotypes.

Affective AI models influence user emotions, which raises ethical concerns around manipulation and trust. There is a higher risk involved when dealing with sensitive applications like suicide detection and well-being assessment. \textbf{Over-reliance on AI} for mental health support is also a potential risk, where \textbf{misinterpretations} or \textbf{insensitive responses} could lead to harm.

The use of personal and emotional data in affective computing raises significant\textbf{ privacy concerns}. Attacks involving data leaks or malicious organizations that sell data could lead to exposure of sensitive private data.

Many sentiment analysis models struggle with fairness when applied across different demographic groups. SenticNet 7 \cite{SenticNet} performs well in sentiment analysis but struggles with nuanced cases like sarcasm and contradictory opinions. Similarly, KnowleNet \cite{knowleNet} relies on external knowledge bases like ConceptNet, which may introduce biases that do not generalize across cultures.

\section{Future Directions}
While the advancement of Affective Computing indeed looks promising, future developments should be rooted in their implications for broader societal challenges. Since the technologies will now find applications in sensitive use cases, they must be able to handle a wide array of situations. Researchers should look to bridge the gap for\textbf{diverse} and \textbf{standardized datasets}. Since data collection is one of the foundational steps in AI development, it is critical that its collection be representative by researchers. When designing Affective Computing solutions, developers need to keep in mind different personas representing users from all ages and cultural groups. This would lead to stronger and more generalizable systems, capable of handling a wide range of use cases. Further, training on larger and more comprehensive datasets alone would improve the generalization and performance of models.

Organizations building Affective Computing-based applications have to realize that these systems are dealing with highly sensitive information. They need, therefore, to have stringent mechanisms in place for \textbf{data privacy}. They should also understand and comply with the \textbf{region-specific data protection laws} that guide the conduct of their operations and ensure ethical and legal compliance.

The rise of Affective Computing in emotion recognition has made applications such as therapy more accessible. This would further extend to a wide group of users whose mental health conditions could be effectively improved, adding to a healthy and balanced society.

\section*{Conclusion}
Affective Computing has gone a long way in improving human-computer interactions by incorporating emotion recognition into AI-driven systems. This survey looked at various works that employ LLMs, multimodal approaches, and affect-aware applications to further the field. These findings confirm that, whereas ChatGPT works excellently on sentiment analyses, personality recognition, and most other applications pertaining to mental health, many challenges are yet to be overcome in critical areas, including dataset diversity, multimodal integration, ethics, and especially real-world usage.

Most of the reviewed studies find that there is a lack of diversity in training datasets, which further causes bias and affects the generalization performance across cultures and languages. Further, the dependency on benchmark datasets alone can hardly represent real-life emotional expressions. This calls for more dataset inclusivity, enhancement of techniques for prompt engineering, and fine-tuning the models for context-sensitive affective tasks.


\begin{thebibliography}{00}
\bibitem{temperature} Zhang, M, Ye, L, Hu, L, Temperature Control Strategy of Passenger Compartment Based on Control Algorithms. Int.J Automot. Technol. 24, 35–43 (2023). \url{https://doi.org/10.1007/s12239-023-0004-y}

\bibitem{hci-rise} Hasyim, H., Bakri, M., ``Advancements in Human-Computer Interaction: A Review of Recent Research,'' Advances: Jurnal Ekonomi \& Bisnis, 2024, https://doi.org/10.60079/ajeb.v2i4.327.

\bibitem{hci}Interaction Design Foundation - IxDF. ``What is Human-Computer Interaction (HCI)?'' Interaction Design Foundation - IxDF,  (accessed Feb. 20, 2025). \url{https://www.interaction-design.org/literature/topics/human-computer-interaction}

\bibitem{handbook} Sidney, D., Rafael A.C., Arvid, K., and Jonathan, G., ``The Oxford Handbook of Affective Computing,'' United States: Oxford University Press, 2014.

\bibitem{affective-roaslind}Picard, R. ``Affective Computing.'' The MIT Press, 1997. \url{https://search.ebscohost.com/login.aspx?direct=true&AuthType=ip,shib&db=e000xna&AN=397&site=ehost-live&scope=site}

\bibitem{biosensor} Kandimalla B. V., Tripathi S. V., Huangxian J., ``Chapter 16 - Biosensors based on immobilization of biomolecules in sol-gel matrices'', Electrochemical Sensors, Biosensors and their Biomedical Applications, 2008, Pages 503-529,
ISBN 9780123737380. \url{https://doi.org/10.1016/B978-012373738-0.50018-0}

\bibitem{biosensor-emotion} Dzedzickis A., Kaklauskas A., Bucinskas V., ``Human Emotion Recognition: Review of Sensors and Methods.'' Sensors (Basel). 2020 Jan 21;20(3):592. doi: 10.3390/s20030592. PMID: 31973140; PMCID: PMC7037130.

\bibitem{journal}``IEEE Transactions on Affective Computing''. \url{https://ieeexplore.ieee.org/xpl/RecentIssue.jsp?punumber=5165369}

\bibitem{openai}OpenAI, ``GPT-4o: A New Multimodal Model,'' OpenAI, May 2024. \url{https://openai.com/index/hello-gpt-4o/?utm_source=chatgpt.com}

\bibitem{transformers} A. A. Shakil and D. M. Ghimire and J. Kalita, ``Using transformers for multimodal emotion recognition: Taxonomies and state of the art review'', Engineering Applications of Artificial Intelligence. \url{https://doi.org/10.1016/j.engappai.2024.108339}

\bibitem{RVISA}W. Lai, H. Xie, G. Xu and Q. Li, ``RVISA: Reasoning and Verification for Implicit Sentiment Analysis,'' in IEEE Transactions on Affective Computing, 2025. \url{https://doi.org/10.1109/TAFFC.2025.3537799}

\bibitem{GPTEval}R. Mao, G. Chen, X. Zhang, F. Guerin, E. Cambria, ``GPTEval: A survey on assessments of ChatGPT and GPT-4'', Proceedings of the 2024 Joint International Conference on Computational Linguistics, Language Resources and Evaluation, 2024. \url{https://doi.org/10.48550/arXiv.2308.12488}

\bibitem{sentiment-analyzer} Z. Wang, Qi. Xie, Y Feng, Z. Ding, Z. Yang, R. Xia, ``Is ChatGPT a Good Sentiment Analyzer? A Preliminary Study'', 2024. \url{https://doi.org/10.48550/arXiv.2304.04339}

\bibitem{personality-recognizer} Y. Ji, W. Wu, H. Zheng, Y. Hu, X. Chen, L. He, ``Is ChatGPT a Good Personality Recognizer? A Preliminary Study'', 2023. \url{https://doi.org/10.48550/arXiv.2307.0395}

\bibitem{nlp-solver} C. Qin, A. Zhang, Z. Zhang,  J. Chen, W. Yasunaga, D. Yang, 2023, ``Is ChatGPT a General-Purpose Natural Language Processing Task Solver?''. \url{https://doi.org/10.48550/arXiv.2302.06476}

\bibitem{gpt-wide-eval} M. M. Amin, R. Mao, E. Cambria, B. W. Schuller, ``A Wide Evaluation of ChatGPT on Affective Computing Tasks'', 2023. \url{https://doi.org/10.48550/arXiv.2308.13911}

\bibitem{sentiment-llm} W. Zhang, Y. Deng, B. Liu, S. Jialin Pan, L. Bing, ``Sentiment Analysis in the Era of Large Language Models: A Reality Check'', 2023. \url{https://doi.org/10.48550/arXiv.2305.15005}

\bibitem{first-gpt} M M. Amin, E. Cambria, B. W. Schuller, ``Will Affective Computing Emerge from Foundation Models and General AI? A First Evaluation on ChatGPT'', 2023. \url{https://doi.org/10.48550/arXiv.2303.03186}

\bibitem{peers}G. Alhussein, M. Alkhodari, A. H. Khandoker and L. J. Hadjileontiadis, ``Novel Speech-Based Emotion Climate Recognition in Peers,'' Conversations Incorporating Affect Dynamics and Temporal Convolutional Neural Networks, in IEEE Access, vol. 13, pp. 16752-16769, 2025. \url{https://doi.org/10.1109/ACCESS.2025.3529125.}

\bibitem{VyaktitvaNirdharan}M. Leekha, S. N. Khan, H. Srinivas, R. R. Shah, J. Shukla, ``VyaktitvaNirdharan: Multimodal Assessment of Personality and Trait Emotional Intelligence,'', in IEEE Transactions on Affective Computing, vol. 15, no. 4, pp. 2139-2153, Oct.-Dec. 2024. \url{https://doi: 10.1109/TAFFC.2024.3404243}

\bibitem{knowleNet}T. Yue, R. Mao, H. Wang, Z. Hu, E. Cambria,
``KnowleNet: Knowledge fusion network for multimodal sarcasm detection'', Information Fusion, Volume 100, 2023. \url{https://www.sciencedirect.com/science/article/pii/S1566253523002373}

\bibitem{sarcasm-news}A. K. Jayaraman, T. E. Trueman, G. Ananthakrishnan, S. Mitra, Q. Liu and E. Cambria, "Sarcasm Detection in News Headlines using Supervised Learning," 2022 International Conference on Artificial Intelligence and Data Engineering (AIDE), Karkala, India, 2022, pp. 288-294, doi: 10.1109/AIDE57180.2022.10060855.

\bibitem{biases}R. Mao, Q. Liu, K. He, W. Li and E. Cambria, ``The Biases of Pre-Trained Language Models: An Empirical Study on Prompt-Based Sentiment Analysis and Emotion Detection,'' in IEEE Transactions on Affective Computing, vol. 14, no. 3, pp. 1743-1753, 1 July-Sept. 2023, doi: 10.1109/TAFFC.2022.3204972.

\bibitem{SenticNet}C. Erik, L. Qian, D. Sergio, X. Frank, K. Kenneth, ``{S}entic{N}et 7: A Commonsense-based Neurosymbolic {AI} Framework for Explainable Sentiment Analysis'', Proceedings of the Thirteenth Language Resources and Evaluation Conference, 2022. \url{https://aclanthology.org/2022.lrec-1.408/}

\bibitem{generativeghosts}
M. R. Morris and J. R. Brubaker, "Generative Ghosts: Anticipating Benefits and Risks of AI Afterlives," arXiv preprint arXiv:2402.01662v2, 2024. \href{https://arxiv.org/abs/2402.01662}{https://arxiv.org/abs/2402.01662}.


\bibitem{grief-chatbot}
A. Xygkou, P. Siriaraya, A. Covaci, H. G. Prigerson, R. Neimeyer, C. S. Ang, and W.-J. She, "The 'Conversation' about Loss: Understanding How Chatbot Technology was Used in Supporting People in Grief," 2023. \url{https://doi.org/10.1145/3544548.3581154}

\bibitem{emma}A. Ghandeharioun, D. McDuff, M. Czerwinski and K. Rowan, ``EMMA: An Emotion-Aware Wellbeing Chatbot'', 8th International Conference on Affective Computing and Intelligent Interaction (ACII), Cambridge, UK, 2019, pp. 1-7. \url{https://doi.org/10.1109/ACII.2019.8925455}

\bibitem{virtual-therapist} L. Ring, T. Bickmore, P. Pedrelli, ``An Affectively Aware Virtual Therapist for Depression Counseling'', 2016. \url{https://api.semanticscholar.org/CorpusID:3911458}

\bibitem{emotake}Y. Gu, Y. Weng, Y. Wang, M. Wang, G. Zhuang, J. Huang, X. Peng, L. Luo, F. Ren, "EmoTake: Exploring Drivers’ Emotion for Takeover Behavior Prediction,", IEEE Transactions on Affective Computing, vol. 15, no. 4, pp. 2112-2127, Oct.-Dec. 2024. \url{https://doi.org/10.1109/TAFFC.2024.3399328}

\bibitem{driver-stress} C. Bustos, N. Elhaouij, A. Sole-Ribalta, J. Borge-Holthoefer, A. Lapedriza, R. Picard, ``Predicting Driver Self-Reported Stress by Analyzing the Road Scene'', 2021. \url{https://doi.org/10.48550/arXiv.2109.1322}

\bibitem{roberta}Y. Liu et al., ``RoBERTa: A Robustly Optimized BERT Pretraining Approach,'' arXiv preprint. \url{https://arxiv.org/abs/1907.11692}

\bibitem{BERT}J. Devlin, M. Chang, K. Lee, and K. Toutanova, ``BERT: Pre-training of Deep Bidirectional Transformers for Language Understanding'', arXiv preprint arXiv:1810.04805, 2018. \url{https://arxiv.org/abs/1810.04805}

\bibitem{MFCC}S. Davis and P. Mermelstein, ``Comparison of parametric representations for monosyllabic word recognition in continuously spoken sentences,''IEEE Transactions on Acoustics, Speech, and Signal Processing, vol. 28, no. 4, pp. 357–366, 1980. {IEEE}. \url{https://doi.org/10.1109/TASSP.1980.1163420}

\bibitem{bigfiveLiverpool} ``The Big Five Personality Traits," University of Liverpool – Prosper Portal'' [Accessed: Apr. 8, 2025]. \url{https://prosper.liverpool.ac.uk/postdoc-resources/reflect/the-big-five/}

\bibitem{conceptnet} "ConceptNet," OpenCyc / Commonsense Computing. \href{https://conceptnet.io/}{https://conceptnet.io/}.

\bibitem{semeval-restaurants} T. Aarsen, "SetFit ABSA SemEval 2014 Restaurants," Hugging Face, 2022. \url{https://huggingface.co/datasets/tomaarsen/setfit-absa-semeval-restaurants}

\bibitem{semeval-laptops} T. Aarsen, "SetFit ABSA SemEval 2014 Laptops," Hugging Face, 2022. \url{https://huggingface.co/datasets/tomaarsen/setfit-absa-semeval-laptops}

\bibitem{KEmoCon} Lee, J., Ko, W., Kim, Y.G. et al. K-EmoCon, a multimodal sensor dataset for continuous emotion recognition in naturalistic conversations. Sci Data 7, 293 (2020). \url{https://doi.org/10.1038/s41597-020-00655-2}

\bibitem{IEMOCAP} Busso, C., Bulut, M., Lee, CC. et al. IEMOCAP: interactive emotional dyadic motion capture database. Lang Resources \& Evaluation 42, 335–359 (2008). \url{https://doi.org/10.1007/s10579-008-9076-6}

\bibitem{SEWA-paper}Kossaifi, J., Walecki, R., Panagakis, Y. et al. SEWA DB: A Rich Database for Audio-Visual Emotion and Sentiment Research in the Wild. arXiv preprint arXiv:1901.02839 (2019). \url{https://arxiv.org/abs/1901.02839}

\bibitem{SEWA-site} "SEWA Project Resources," SEWA Project, . [Accessed: Apr. 8, 2025]. \url{https://www.sewaproject.eu/resources}

\bibitem{superglue} "SuperGLUE Benchmark," . [Accessed: Apr. 8, 2025]. \url{https://super.gluebenchmark.com/}

\bibitem{xsum} "XSum Dataset," Hugging Face – EdinburghNLP, . [Accessed: Apr. 8, 2025]. \url{https://huggingface.co/datasets/EdinburghNLP/xsum}

\bibitem{wikibio}Lebret, R., Grangier, D., and Auli, M. "Neural Text Generation from Structured Data with Application to the Biography Domain," In *Proceedings of EMNLP 2016*.

\bibitem{mmlu} "Massive Multitask Language Understanding (MMLU) Dataset," Hugging Face, . [Accessed: Apr. 8, 2025]. \url{https://huggingface.co/datasets/cais/mmlu}

\bibitem{gsm8k} "GSM8K Dataset," Hugging Face – OpenAI, . [Accessed: Apr. 8, 2025]. \url{https://huggingface.co/datasets/openai/gsm8k}

\bibitem{pubmedqa} "PubMedQA Dataset," GitHub, . [Accessed: Apr. 8, 2025]. \url{https://github.com/pubmedqa/pubmedqa}

\bibitem{medqa} Jin, D. et al. "What Disease does this Patient Have? A Large-scale Open Domain Question Answering Dataset from Medical Exams," *arXiv preprint arXiv:2009.13081*, 2020.

\bibitem{codexglue} Lu, S. et al. "CodeXGLUE: A Machine Learning Benchmark Dataset for Code Understanding and Generation," *CoRR*, vol. abs/2102.04664, 2021.

\bibitem{stereoset} Nadeem, M., Bethke, A., and Reddy, S. "StereoSet: Measuring stereotypical bias in pretrained language models," *arXiv preprint arXiv:2004.09456*, 2020.

\bibitem{crows}Nangia, N. et al. "CrowS-Pairs: A Challenge Dataset for Measuring Social Biases in Masked Language Models," *Proceedings of EMNLP 2020*. \url{https://aclanthology.org/2020.emnlp-main.154/}

\bibitem{sst2} "Stanford Sentiment Treebank (SST-2)," Hugging Face – StanfordNLP, . [Accessed: Apr. 8, 2025]. \url{https://huggingface.co/datasets/stanfordnlp/sst2}

\bibitem{emotionx} Shmueli, B. and Ku, L.-W. "SocialNLP EmotionX 2019 Challenge Overview: Predicting Emotions in Spoken Dialogues and Chats," *arXiv preprint arXiv:1909.07734*, 2019.

\bibitem{emotakedataset} ``EmoTake''. \url{https://github.com/yibingweng/EmoTake}

\bibitem{emotion-cause-dataset}L. Gui, D. Wu, R. Xu, Q. Lu, and Y. Zhou, "Event-Driven Emotion Cause Extraction with Corpus Construction," in \textit{Proceedings of the 2016 Conference on Empirical Methods in Natural Language Processing (EMNLP)}, Austin, Texas, Nov. 2016, pp. 1639–1649. \url{https://aclanthology.org/D16-1170/}

\bibitem{essays-dataset}J. W. Pennebaker and L. A. King, "Linguistic styles: Language use as an individual difference," \textit{Journal of Personality and Social Psychology}, vol. 77, no. 6, pp. 1296–1312, 1999. \url{https://doi.org/10.1037/0022-3514.77.6.1296}

\bibitem{pan-dataset}F. Rangel, P. Rosso, M. Koppel, E. Stamatatos, and G. Inches, "Overview of the 3rd Author Profiling Task at PAN 2015," in \textit{Working Notes of CLEF 2015 Conference and Labs of the Evaluation Forum}, 2015. \url{https://pan.webis.de/clef15/pan15-web/author-profiling.html}

\bibitem{sarcasm-twitter}Y. Cai, H. Cai, X. Wan, ``Multi-modal sarcasm detection in Twitter with hierarchical fusion model'', Proceedings of the 57th Annual Meeting of the Association
for Computational Linguistics, 2019, pp. 2506–2515.

\bibitem{sarcasm-twitred}K. Maity, P. Jha, S. Saha, P. Bhattacharyya, ``A multitask framework for sentiment, emotion and sarcasm aware cyberbullying detection from multi-modal code-
mixed memes'', Proceedings of the 45th International ACM SIGIR Conference on Research and Development in Information Retrieval, 2022, pp. 1739–1749.

\bibitem{csqa} T. Sap, R. Le Bras, E. Bhagavatula, and Y. Choi, "CommonsenseQA: A Question Answering Challenge Targeting Commonsense Knowledge," in *Proc. NAACL*, 2019. \url{https://aclanthology.org/N19-1421/}

\bibitem{copa} G. B. Gordon, R. Mihalcea, and J. Louis, "Choice of Plausible Alternatives: An Evaluation of Commonsense Causal Reasoning," in *Proc. SemEval*, 2012. \url{https://aclanthology.org/S12-1051/}

\bibitem{rte} D. Dagan, O. Glickman, and B. Magnini, "The PASCAL Recognising Textual Entailment Challenge," in *Machine Learning Challenges*, 2006. \url{https://aclanthology.org/W05-1201/}

\bibitem{cb} A. De Marneffe, C. Manning, et al., "The CommitmentBank: Investigating projection in naturally occurring discourse," in *Proc. Sinn und Bedeutung*, 2019. \url{https://github.com/mcdm/CommitmentBank}

\bibitem{boolq} Y. Clark et al., "BoolQ: Exploring the Surprising Difficulty of Natural Yes/No Questions," in *Proc. NAACL*, 2019. \url{https://aclanthology.org/N19-1300/}

\bibitem{mutual} H. Cui et al., "CoMPM: A Comprehensive Multi-turn Dialogue Reasoning Benchmark," in *Proc. ACL*, 2020. \url{https://aclanthology.org/2020.acl-main.743/}

\bibitem{samsum} P. Gliwa, I. Mochol, M. Biesek, and A. Wawer, "SAMSum Corpus: A Human-annotated Dialogue Dataset for Abstractive Summarization," in *Proc. EMNLP*, 2019. \url{https://aclanthology.org/D19-5409/}

\bibitem{conll03} E. Tjong Kim Sang and F. De Meulder, "Introduction to the CoNLL-2003 Shared Task: Language-Independent Named Entity Recognition," in *Proc. CoNLL*, 2003. \url{https://aclanthology.org/W03-0419/}

\bibitem{twitter140} A. Go, R. Bhayani, and L. Huang, ``Twitter Sentiment Classification using
Distant Supervision'', CS224N project report, Stanford, vol. 1, no. 12, p. 2009, 2009.

\bibitem{semeval-2015}K. Cortis, A. Freitas, T. Daudert, M. Huerlimann, M. Zarrouk, S. Handschuh, and B. Davis, ``SemEval-2017 Task 5: Fine-Grained Sentiment Analysis on Financial Microblogs and News'', in Proceedings of SemEval-2017, 2017, pp. 519–535.

\bibitem{wassa-2017}S. Mohammad and F. Bravo-Marquez, ``WASSA-2017 Shared Task on Emotion Intensity'', in Proceedings of WASSA 2017, 2017, pp. 34–49.

\bibitem{teenage-reddit} V. Desu, N. Komati, S. Lingamaneni, and F. Shaik, “Suicide and Depression Detection in Social Media Forums,” in Smart Intelligent Computing and Applications, Volume 2. Singapore, Singapore: Springer Nature Singapore, 2022, pp. 263–270.

\bibitem{toxic} cjadams, Jeffrey Sorensen, Julia Elliott, Lucas Dixon, Mark McDonald, nithum, and Will Cukierski. Toxic Comment Classification Challenge. https://kaggle.com/competitions/jigsaw-toxic-comment-classification-challenge, 2017. Kaggle.

\bibitem{well-being} A. Rastogi, Q. Liu, and E. Cambria, ``Stress Detection from Social Media Articles: New Dataset Benchmark and Analytical Study,'' in IJCNN, 2022, pp. 1–8.

\bibitem{tedtalks-kaggle} "Social Media Interactions on TED Talks Dataset," Kaggle, 2022. . [Accessed: Apr. 8, 2025]. \url{https://www.kaggle.com/datasets/thedevastator/social-media-interactions-on-tedtalks-dataset}

\bibitem{first-impressions} V. Ponce-L ´opez, B. Chen, M. Oliu, C. Corneanu, A. Clap ´es, I. Guyon, X. Bar ´o, H. J. Escalante, and S. Escalera, “Chalearn lap 2016: First Round Challenge on First Impressions - Dataset and Results,” in ECCV, Cham, Switzerland, 2016, pp. 400–418.

\bibitem{onion-huff} R. Misra and P. Arora, “Sarcasm Detection using News Headlines Dataset,” AI Open, vol. 4, pp. 13–18, 2023.

\bibitem{rotten-tomatoes}B. Pang and L. Lee, “A Sentimental Education: Sentiment Analysis Using Subjectivity Summarization Based on Minimum Cuts,” in Proceedings of ACL, 2004, pp. 271–278.

\bibitem{imdb}J. J. Jindal and M. Liu, "Identifying comparative sentences in text documents," in \textit{Proceedings of the 49th Annual Meeting of the Association for Computational Linguistics: Human Language Technologies}, 2011, pp. 109–117. \url{https://aclanthology.org/P11-1015/}

\bibitem{yelp-2-5}X. Zhang, J. Zhao, and Y. LeCun, "Character-level Convolutional Networks for Text Classification," in \textit{Advances in Neural Information Processing Systems (NeurIPS)}, vol. 28, 2015. \url{https://proceedings.neurips.cc/paper_files/paper/2015/file/250cf8b51c773f3f8dc8b4be867a9a02-Paper.pdf}

\bibitem{MR}B. Pang and L. Lee, "Seeing Stars: Exploiting Class Relationships for Sentiment Categorization with Respect to Rating Scales," in \textit{Proceedings of the 43rd Annual Meeting of the Association for Computational Linguistics (ACL'05)}, Ann Arbor, Michigan, Jun. 2005, pp. 115–124. \url{https://aclanthology.org/P05-1015/}

\bibitem{twitter-sentiment-llm}S. Rosenthal, N. Farra, and P. Nakov, "SemEval-2017 Task 4: Sentiment Analysis in Twitter," in \textit{Proceedings of the 11th International Workshop on Semantic Evaluation (SemEval-2017)}, Vancouver, Canada, Aug. 2017, pp. 502–518. \url{https://aclanthology.org/S17-2088/}

\bibitem{sst5}R. Socher et al., "Recursive Deep Models for Semantic Compositionality Over a Sentiment Treebank," in \textit{Proceedings of the 2013 Conference on Empirical Methods in Natural Language Processing (EMNLP)}, Seattle, Washington, USA, Oct. 2013, pp. 1631–1642. \url{https://aclanthology.org/D13-1170/}

\bibitem{hateeval}V. Basile et al., "SemEval-2019 Task 5: Multilingual Detection of Hate Speech Against Immigrants and Women in Twitter," in \textit{Proceedings of the 13th International Workshop on Semantic Evaluation (SemEval-2019)}, Minneapolis, Minnesota, USA, Jun. 2019, pp. 54–63. \url{https://aclanthology.org/S19-2007/}

\bibitem{irony18}C. Van Hee, E. Lefever, and V. Hoste, "SemEval-2018 Task 3: Irony Detection in English Tweets," in \textit{Proceedings of the 12th International Workshop on Semantic Evaluation (SemEval-2018)}, New Orleans, Louisiana, Jun. 2018, pp. 39–50. \url{https://aclanthology.org/S18-1005/}

\bibitem{offenseval}M. Zampieri, S. Malmasi, P. Nakov, S. Rosenthal, N. Farra, and R. Kumar, "SemEval-2019 Task 6: Identifying and Categorizing Offensive Language in Social Media (OffensEval)," in \textit{Proceedings of the 13th International Workshop on Semantic Evaluation (SemEval-2019)}, Minneapolis, Minnesota, Jun. 2019, pp. 75–86. \url{https://aclanthology.org/S19-2010/}

\bibitem{stance16}S. Mohammad, S. Kiritchenko, P. Sobhani, X. Zhu, and C. Cherry, "SemEval-2016 Task 6: Detecting Stance in Tweets," in \textit{Proceedings of the 10th International Workshop on Semantic Evaluation (SemEval-2016)}, San Diego, California, Jun. 2016, pp. 31–41. \url{https://aclanthology.org/S16-1003/}

\bibitem{cs19}A. Panchenko, A. Bondarenko, M. Franzek, M. Hagen, and C. Biemann, "Categorizing Comparative Sentences," in \textit{Proceedings of the 6th Workshop on Argument Mining}, Florence, Italy, Aug. 2019, pp. 136–145. \url{https://aclanthology.org/W19-4516/}

\bibitem{emotion20} F. Barbieri, J. Camacho-Collados, L. Espinosa Anke, and L. Neves, "TweetEval: Unified Benchmark and Comparative Evaluation for Tweet Classification," in \textit{Findings of the Association for Computational Linguistics: EMNLP 2020}, Online, Nov. 2020, pp. 1644–1650. \url{https://aclanthology.org/2020.findings-emnlp.148/}

\bibitem{dep-detection} V. Desu et al., “Suicide and Depression Detection in Social Media Forums,” Smart Intelligent Computing and Applications, Volume 2, Springer Nature Singapore, Singapore, 2022, pp. 263–270.

\bibitem{amazon-reviews} J. Blitzer, M. Dredze, and F. Pereira, “Biographies, bollywood, boom-boxes and blenders: Domain adaptation for sentiment classification,” in Proc. 45th Annu. Meeting Assoc. Comput. Linguistics, 2007, pp. 440–447.

\bibitem{customer-reviews} Hu, M. and Liu, B., ``Mining and summarizing customer reviews'', 2004, SIGKDD, pages 168–177.

\bibitem{sanders} "Sanders Twitter Sentiment Corpus," GitHub, 2011. . [Accessed: Apr. 8, 2025]. \url{https://github.com/zfz/twitter_corpus}

\bibitem{sts}Saif, H., Fernandez, M., He, Y., and Alani, H., "Evaluation datasets for twitter sentiment analysis: a survey and a new dataset, the sts-gold", In International Conference of the Italian Association for Artificial Intelligence, 2013.

\bibitem{affectiveroad-dataset} "AffectiveROAD Dataset," MIT Media Lab – Affective Computing Group, 2018. . [Accessed: Apr. 8, 2025]. \url{https://www.media.mit.edu/groups/affective-computing/data/}

\bibitem{speech-transcript}S. A. Burkhardt, "Emotional Prosody Speech and Transcripts," Linguistic Data Consortium, LDC2002S28, 2002. \url{https://catalog.ldc.upenn.edu/LDC2002S28}

\bibitem{amfed} D. McDuff, R. El Kaliouby, T. Senechal, M. Amr, J. F. Cohn, and R. Picard, "Affectiva-MIT Facial Expression Dataset (AM-FED): Naturalistic and Spontaneous Facial Expressions Collected In-the-Wild," MIT Media Lab and Affectiva Inc.  [Accessed: Apr. 8, 2025]. \url{https://www.affectiva.com/}
\end{thebibliography}
\end{document}